\documentclass[12pt,letterpaper]{article}
\usepackage{jheppub}
\usepackage{verbatim}
\usepackage{graphicx}
\usepackage{epstopdf}
\usepackage{amsthm}
\usepackage[T1]{fontenc}
\usepackage{lmodern}
\newcommand{\hiddensubsection}[1]{%
	\refstepcounter{subsection}%
	\subsection*{\thesubsection\quad #1}%
}
\newcommand{\Tr}{\text{Tr}}

\title{Entanglement and Charge Transfer}

\author[]{Tom Shachar}
\affiliation{Instituto Balseiro, Centro Atómico Bariloche, S.C. Bariloche 8400, Río Negro, Argentina}
\emailAdd{tom.shachar@ib.edu.ar}

\abstract{
	We compute the entropic order parameter introduced in (1905.10487) using the replica trick for two disjoint intervals. For a global symmetry, this quantity measures the mutual information carried by charge-anticharge intertwiners. We formulate the problem in terms of averages over twisted partition functions on replica Riemann surfaces, and obtain exact solutions for free fermions and compact bosons, for $U(1)$ symmetry and its discrete $\mathbb Z_N$ subgroups. We show that further averaging over both families of non-contractible cycles gives access to an additional quantity, which is the loss of Holevo information associated with complementary-charge measurements that occurs upon removing the intertwiners. For the boson, this quantity vanishes, whereas for Dirac fermions it equals the difference between the mutual information of the fermion-parity-gauged and ungauged theories, independently of the choice of even $\mathbb Z_N$ subgroup. Finally, we extend the analysis to non-Abelian symmetries using saddle expansions, deriving universal behavior in the nearly touching intervals limit and for a large number of fermions $N_F\gg1$.
}

\begin{document}
	\maketitle
	\section{Introduction}
	In the last few decades, entanglement has played a pivotal role in advancing our understanding of QFT. In particular, the algebraic formulation of entanglement in QFT has provided a unique vantage point that brings to view some subtle physical questions. A key observation along that line, is that in some cases an ambiguity arises where more than one non-preferable choice of algebra is possible for a given region. From the original algebraic perspective, this is known as Haag-duality violation, and it divides between global symmetries \cite{Doplicher1969,Doplicher1969a,Doplicher1974}, and gauge theories \cite{Buchholz1982}. The two algebras generally prescribe different entanglement content to the same region, with the difference characterized by a relative entropy satisfying special properties \cite{Longo2018,Xu2020,Casini2020,Casini2021a,Casini2022,PhysRevA.103.012211,Furuya2020,Hollands2021}. We shall follow the terminology of \cite{Casini2020} and call it the entropic order parameter, or for brevity, EOP from hereafter. The EOP has been studied primarily using algebraic techniques, and while it exhibits some universal behavior in certain regimes, complete solutions are analytically known only for very few models \cite{PhysRevD.98.125008,Casini2020,Benedetti:2026nlt}. As we explain in more detail below, for a theory with a global symmetry, there is a very simple physical interpretation. When one wishes to restrict to a subalgebra of neutral operators under the symmetry, there are potentially charge-anticharge pairs that are neutral in the entire region but not in each disconnected subregion individually. These pairs, also known as intertwiners, transfer charge between the two disconnected subregions and cannot be locally generated by neutral operators in each one individually. The EOP therefore measures the contribution of those charge-transfer operators to the mutual information between the two intervals.

	In this paper we set to study the EOP using the replica trick and thereby achieving two goals. First, we establish connection with other quantum-information quantities and symmetry related measures of entanglement. Secondly, we obtain new exact results and approximation schemes by direct path-integral computations.
	The main findings include an operational interpretation of the EOP in terms of Holevo information related to charge measurements in the complementary region, and more importantly a path-integral method of computing the change of this quantity under the removal of the intertwiners. For free Dirac fermions, we find that this change is precisely equal to the difference between the mutual information of the fermionic $\mathbb Z^F_2$-gauged theory with that of the ungauged fixed spin-structure theory. Additionally, we present explicit expressions for the two-interval EOP in free massless fermionic theories in two-dimensions, and likewise for the compact boson at general compactification radius. We analyze both the $U\!\left(1\right)$ symmetry and its $\mathbb Z_N$ subgroups, and for the boson treat both the field-shifting and winding symmetries. The results are primarily exact at integer R\'enyi index, and when possible, we also discuss the continuation to the von-Neumann limit. Subsequently, with the aim of going beyond the exactly solvable Abelian examples, we explore two approximation methods for a general symmetry group. We do so for a large number of fermions, and for the nearly touching interval limit, where we reproduce the desired universal behavior for the latter.
	
	The manuscript is organized as follows: In section \ref{section-EOP} we introduce the density matrix and replica formulations of the EOP, as well as its relation to complementary charge measurements and Holevo information. In section \ref{section-exactresults} we consider exactly solvable examples, beginning with free fermions and subsequently considering the compact boson in subsection \ref{sec:compact_boson} and finite $\mathbb{Z}_N$ symmetry group in subsection \ref{sec:Zn_projection}. In section \ref{section-nonAbelian}, we present two approximation methods for more general non-Abelian symmetries. Finally, we conclude with a discussion.
	
	\section{Entropic Order Parameter}
	\label{section-EOP}
\begin{figure}[htbp]
		\centering
		\includegraphics[width=0.75\textwidth]{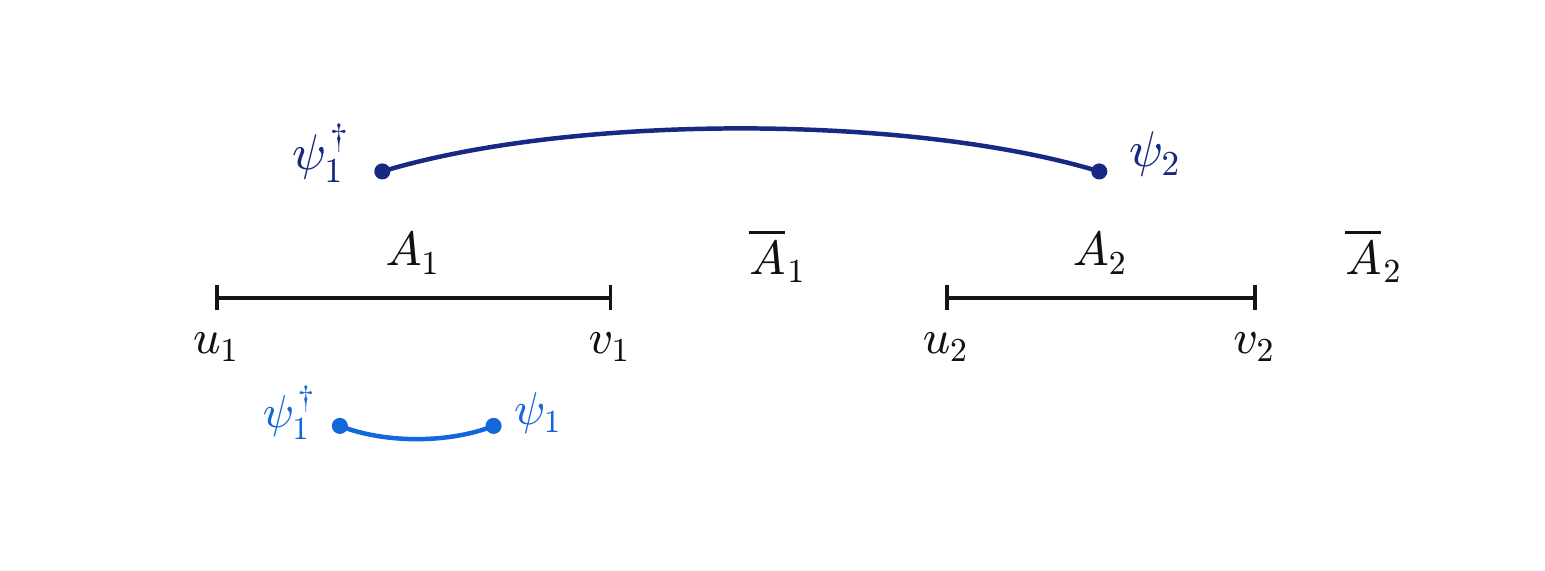}
		\caption{On the plane, the entanglement region consists of two intervals $A=A_1\cup A_2$. The charge-anticharge operators linking the two intervals, also known as intertwiners, are part of the full neutral algebra $\mathcal{N}(A)$ (top), but not the additive one, $\mathcal{N}(A_{1})\otimes\mathcal{N}(A_{2})$ (bottom).}
		\label{fig1}
	\end{figure}

	Perhaps the simplest example of this algebra ambiguity and its physical significance is displayed by the 2d free massless fermion and its $U\!\left(1\right)$ symmetry. However, the replica-trick formulation provided below easily generalizes to a general Lie, or discrete group. Consider the reduced density matrix $\rho_{A}=\Tr_{\bar{A}}\left(\left|\Omega\right\rangle \!\left\langle \Omega\right|\right)$ of the vacuum state restricted to the union of two disjoint spatial intervals $A=A_1\cup A_2$, as depicted in Fig.~\ref{fig1}.
	 \\
	The global $U\!\left(1\right)$ invariance of the vacuum implies that  $\left[\rho_{A},Q_{A}\right]=0$, where the total charge $Q_A$ is the sum of charges in each interval, $Q_A=Q_1+Q_2$. Therefore, the reduced vacuum state is block diagonal with respect to the total charge,
	\begin{equation} \label{projectors}
		\rho_{A}=\underset{q_{A}}{\sum}\Pi_{q_{A}}\rho_{A}\Pi_{q_{A}},\qquad \Pi_{q_A}=\int_{-\pi}^{\pi}\frac{d\theta}{2\pi}e^{i\theta\left(Q_A-q_A\right)}.
	\end{equation}
	Where $\Pi_{q_{A}}$ is the projector to a $Q_A$ charge sector. In terms of operator algebras, this means that $\rho_A$ acts only non-trivially on the subalgebra of neutral operators, $\mathcal{N}(A)$. Indeed, for a charged operator, $\Tr_A\left(\rho_{A}\mathcal{O}\right)=0$. Nevertheless, in general ${[\rho_A,Q_1]\ne0}$. In each sector (with $q_A=q_1+q_2$ fixed) there can be transfer of charge between the two intervals \cite{Casini2009,LONGO2010,DiGiulio2023,Mintchev2025}. Such fluctuations are probed by bi-local operators such as $\psi^{\dagger}\left(x\in A_{1}\right)\psi\left(y\in A_{2}\right)$. Those operator are usually called intertwiners, and while they belong to the maximal neutral algebra $\mathcal{N}(A)$, they do not belong in the additive neutral algebra,  $\mathcal{N}(A_{1})\otimes\mathcal{N}(A_{2})$. Namely, they cannot be generated locally from neutral operators that are localized in one of the intervals, and in this sense they are regarded as non-local with respect to $A$. The maximal and additive neutral subalgebras generally give rise to different entanglement entropies. The additional contribution associated with the charge-transfer intertwiners is quantified by the entropic order parameter.

	In order to define the EOP, we may restrict the state $\rho_A$ into $\mathcal{N}(A_{1})\otimes\mathcal{N}(A_{2})$ by performing the following projection, enforcing neutrality in both intervals,\footnote{The group average \eqref{rhotwidle} is the pullback of the conditional expectation ${\mathcal{E}_{A}:\mathcal{N}\left(A\right)\rightarrow\mathcal{N}\left(A_{1}\right)\otimes\mathcal{N}\left(A_{2}\right)}$. The interested reader may refer to e.g. \cite{Casini2020,Furuya2020} for more information.}
	\begin{equation} \label{rhotwidle}
		\tilde{\rho}_{A}=\underset{-\pi}{\overset{\pi}{\int}}\,\underset{-\pi}{\overset{\pi}{\int}}\frac{d\gamma_1}{2\pi}\frac{d\gamma_2}{2\pi}e^{-i\gamma_1 Q_{1}-i\gamma_2 Q_{2}}\rho_{A}e^{i\gamma_1 Q_{1}+i\gamma_2 Q_{2}}=\underset{q_{1},q_{2}}{\sum}\Pi_{q_{1}}\Pi_{q_{2}}\rho_{A}\Pi_{q_{2}}\Pi_{q_{1}},
	\end{equation}
	Where $\Pi_{q_{1}}$ projects to $Q_1$ and $\Pi_{q_{2}}$ to $Q_2$. It is simple to check that ${\left[\tilde{\rho}_{A},Q_{1}\right]=\left[\tilde{\rho}_{A},Q_{2}\right]=0}$. More so, we may lift the property $\left[\rho_{A},Q_{1}\right]=-\left[\rho_{A},Q_{2}\right]$ into $e^{-i\gamma Q_{2}}\rho_{A}e^{i\gamma Q_{2}}=e^{i\gamma Q_{1}}\rho_{A}e^{-i\gamma Q_{1}}$, from which \eqref{rhotwidle} simplifies,
	\begin{equation} \label{dephasedA}
		\tilde{\rho}_{A}=\underset{-\pi}{\overset{\pi}{\int}}\frac{d\gamma}{2\pi}e^{-i\gamma Q_{1}}\rho_{A}e^{i\gamma Q_{1}}=\underset{q_{1}}{\sum}\Pi_{q_{1}}\rho_{A}\Pi_{q_{1}}.
	\end{equation}
	Indeed, to project out the intertwiners it suffices to enforce neutrality only in a single interval. The EOP is defined as the relative entropy between the vacuum restricted to these two neutral subalgebras $S\left(\rho_{A}||\tilde{\rho}_{A}\right)$. It is then simple to show,
	\begin{equation} \label{EOP-Sdiff}
		S\left(\rho_{A}||\tilde{\rho}_{A}\right)=S\left(\tilde{\rho}_{A}\right)-S_{A}\left(\rho_{A}\right).
	\end{equation}
	Additionally, we notice that for a $U\!\left(1\right)$ symmetric state such as the vacuum, for a \textit{single} interval $\tilde{\rho}_{A_{1}}=\rho_{A_{1}},\;\tilde{\rho}_{A_{2}}=\rho_{A_{2}}$. Therefore,
	\begin{equation} \label{neutral-MI}
		S\left(\rho_{A}||\tilde{\rho}_{A}\right)=I_{A}-\tilde{I}_{A},
	\end{equation}
	with $I_{A},\tilde{I}_{A}$ the mutual informations of the two intervals, evaluated before and after the group-averaging respectively.

	In order to use the replica trick, we need to compute the moments of \eqref{dephasedA},\footnote{We assume the vacuum is normalized such that $\Tr_A(\rho_A)=1$.}
	\begin{equation} \label{charged-moment_def}
		\Tr_A\left(\tilde{\rho}_{A}^{n}\right)=\int\frac{d^{n}\gamma}{\left(2\pi\right)^{n}}Z_{n}\left(\gamma_{i}\right),
	\end{equation}
	\begin{equation} \label{charged-moment_def2}
		Z_{n}\left(\gamma_{i}\right)=\Tr_A\left(\rho_{A}e^{i\left(\gamma_{1}-\gamma_{2}\right)Q_{1}}\rho_{A}e^{i\left(\gamma_{2}-\gamma_{3}\right)Q_{1}}\dots\rho_{A}e^{i\left(\gamma_{n}-\gamma_{1}\right)Q_{1}}\right).
	\end{equation}
	As the dependence is only in the phase differences  $\gamma_{i}-\gamma_{i+1}$, it is clear that there is one redundant variable that can be integrated away. However before doing so, it is instructive to notice that the circle $\gamma_{1}=\gamma_{2}=\dots=\gamma_{n}$ is a degenerate global maximum of $\left|Z_{n}\right|$,
	\begin{equation} \label{charged-moment_n}
		\left|Z_{n}\left(\gamma_{i}\right)\right|\leq Z_{n}\left(0\right)=Z_{n}\left(\gamma_{1}=\gamma_{2}=\dots=\gamma_{n}\right).
	\end{equation}
	This can be seen from the Hölder inequality, $\left\Vert \underset{i}{\overset{n}{\Pi}}\left(U_{i}\rho U_{i}^{\dagger}\right)\right\Vert _{1}\leq\underset{i}{\overset{n}{\Pi}}\left\Vert U_{i}\rho U_{i}^{\dagger}\right\Vert _{n}=\underset{i}{\overset{n}{\Pi}}\left\Vert \rho\right\Vert _{n}$.\footnote{$\left\Vert \cdot\right\Vert _{n}$ denotes the Schatten $n$-norm. Since $\rho$ is positive and Hermitian we have $\left\Vert \rho\right\Vert _{n}=Tr\left(\rho_{A}^{n}\right)^{\frac{1}{n}}$.} Finally, we change integration variables to $\alpha_{i}=\gamma_{i}-\gamma_{i+1}$, and choose the last $n$'th variable to be $\gamma_n$. The last variable in \eqref{charged-moment_def2} can be expressed as $\gamma_n-\gamma_1=-\sum^{n-1}_i \alpha_i$. Thus, \eqref{charged-moment_def} reduces to,\footnote{The difference or sum of the phases $\gamma_i$ should always be taken modulo $2\pi$, as they should be properly treated as a $U\left( 1\right) $ Haar measure.}
	\begin{equation} \label{charged-moment_n-1}
		\Tr_A\left(\tilde{\rho}_{A}^{n}\right)=\int\frac{d^{n-1}\alpha}{\left(2\pi\right)^{n-1}}Z_{n}\left(\alpha_{i}\right).
	\end{equation}
	The entanglement entropy is retrieved by the standard method of taking the limit,
	\begin{equation} \label{Reyni_def}
		\tilde{S}_{A}=-\Tr_A\left(\tilde{\rho}_{A}\log\tilde{\rho}_{A}\right)=\underset{n\rightarrow1}{\lim}\,\tilde{S}_{A}^{\left(n\right)},\qquad\tilde{S}_{A}^{\left(n\right)}=\frac{1}{1-n}\log \Tr_A\left(\tilde{\rho}_{A}^{n}\right).
	\end{equation}
	And in particular due to \eqref{EOP-Sdiff}, the EOP from the ratio,
	\begin{equation} \label{EOP-from-ratio}
		S\left(\rho_{A}||\tilde{\rho}_{A}\right)=\underset{n\rightarrow1}{\lim}\,\frac{1}{1-n}\log\frac{\Tr_A\left(\tilde{\rho}_{A}^{n}\right)}{\Tr_A\left(\rho_{A}^{n}\right)}.
	\end{equation}
	In various different contexts, $Z_n(\alpha_i)$ are usually called charged moments. The computation of such moments also arises naturally in the study of entanglement asymmetry \cite{Ares2023,Fossati2024,Benini:2024xjv} and a variation thereof in symmetry-resolved entanglement entropy \cite{PhysRevLett.120.200602,PhysRevB.98.041106,Ares2022,Gaur2024}. While the two quantities share the same mathematical structure, their physical interpretations differ. Entanglement asymmetry is used to quantify symmetry breaking in a reduced state, whereas the EOP can be nonzero even in a globally symmetric state such as the vacuum. In the case at hand, it measures the contribution of non-local charge-transfer correlations between the disconnected components of the region.
	
	\subsection{Replica Surface and Symmetry Defects}
	\label{subsection-replica-and-symmdef}
\begin{figure}[htbp]
		\centering
		\includegraphics[width=0.6\textwidth]{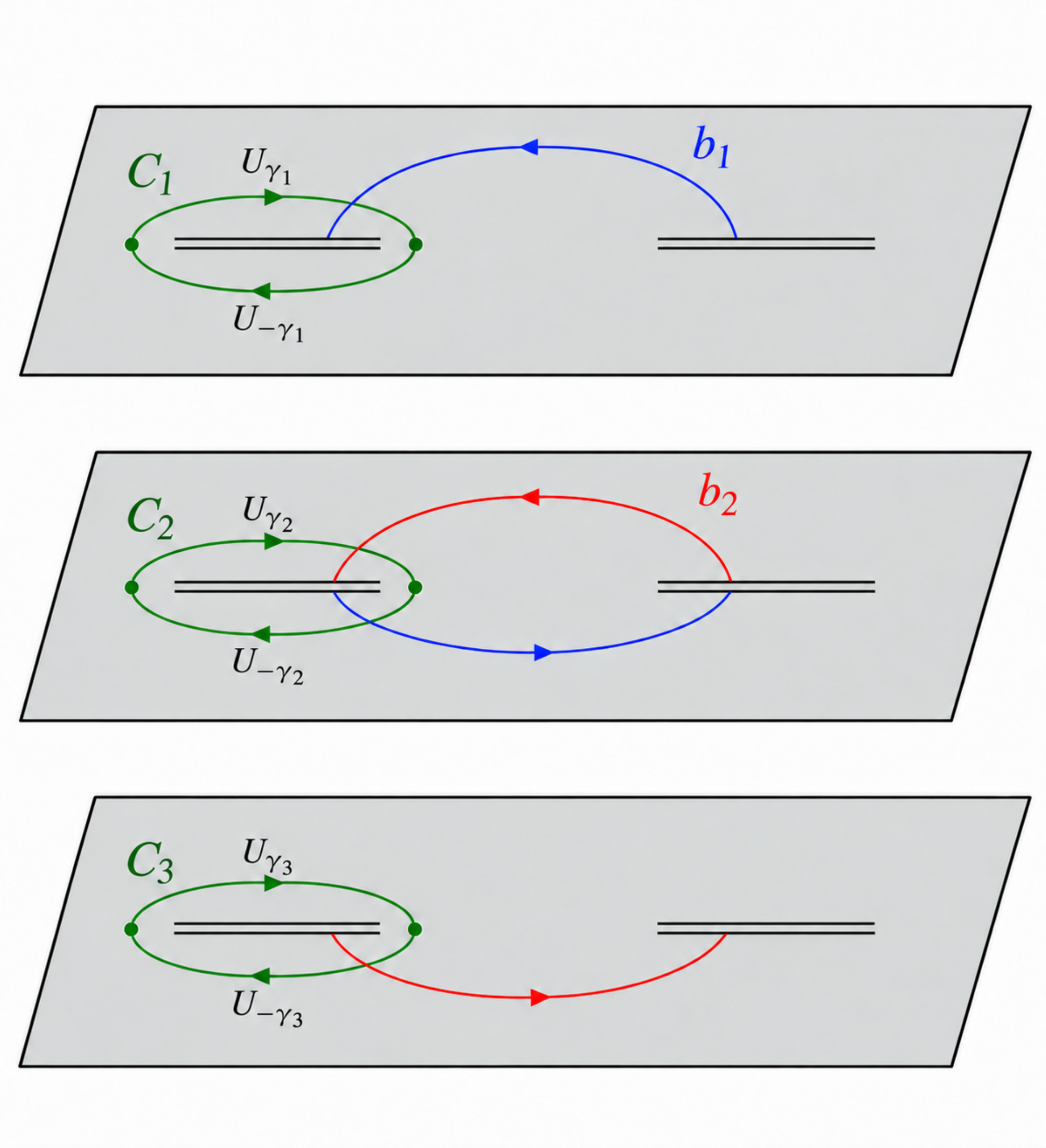}
		\caption{The replica manifold and the symmetry defects for $n=3$. The green contours $C_j$ surround the left branch cut and carry the symmetry defects $U_{\gamma_{j}}\left[C_{j}\right]$, whose upper and lower segments $C_{j}^{+},C_{j}^{-}$ correspond to opposite symmetry operators. The blue and red curves represent the canonical $b_1$ and $b_2$-cycles respectively, with each becoming a closed curve once gluing the branch cuts.}
		\label{fig2}
	\end{figure}

	The replica partition function $Z_n$ is defined on the genus $g=n-1$ Riemann surface $\mathcal{R}_n$, obtained by cyclically gluing $n$ copies of the plane along the two intervals. The charged moments \eqref{charged-moment_def2} evaluate those partition functions with the additional insertions of $U\!\left(1\right)$ symmetry defects \cite{Fossati2024,Bruno2024,Benini:2024xjv,Benedetti:2026nlt}. More precisely, we define the $U\!\left(1\right)$ topological line defect by $U_{\gamma}\left[C\right]=e^{i\gamma\int_{C}\ast J}$. The contours that arise naturally in \eqref{charged-moment_def2} are the closed contours $C_j$ surrounding the left branch cut on each sheet, as depicted in Fig.~\ref{fig2}. Each contour can be split as  $U_{\gamma_{j}}\left[C_{j}\right]=U_{\gamma_{j}}\left[C_{j}^{+}\right]U_{\gamma_{j}}^{\dagger}\left[C_{j}^{-}\right]$, and the two factors can be identified with the unitary operators inserted in the charged moments \eqref{charged-moment_def2}, up to subtleties that we discuss momentarily.
	The contours obey the homology class relation $\sum_{j=1}^{n}[C_j]=0$, which leaves $g=n-1$ independent cycles. To relate these cycles to the $n-1$ independent phases $\alpha_{i}=\gamma_{i}-\gamma_{i+1}$ used in \eqref{charged-moment_n-1}, we follow \cite{Calabrese2009,Calabrese2011,Coser2016} and define a canonical homology basis $\{a_i,b_i\}_{i=1}^{n-1}$ that is adapted to the two branch cuts.\footnote{A genus-$g$ Riemann surface has $2g$ independent non-contractible cycles. The canonical homology basis consists of cycles $\{a_i,b_i\}_{i=1}^{g}$ satisfying $a_i\circ a_j=b_i\circ b_j=0,\; a_i\circ b_j=\delta_{ij},$ where $\circ$ denotes the oriented intersection number.}
	In order to construct the canonical a-cycles, we notice that by definition the intersection number is,
	\begin{equation}
		[C_j]\circ[b_i]
		=
		\delta_{ji}-\delta_{j,i+1}.
	\end{equation}
	We should therefore define,
	\begin{equation}
		[a_k]=\sum_{j=1}^{k}[C_j],
		\qquad k=1,\ldots,n-1,
	\end{equation}
	Or equivalently,
	\begin{equation} \label{homology-basis-transformation}
		[C_j]=[a_j]-[a_{j-1}],
		\qquad a_0=a_n\equiv0.
	\end{equation}
	Since the contours $C_j$ are chosen to be mutually non-intersecting, the topological nature of the defects and their Abelian fusion rules allow us to lift the homology relations \eqref{homology-basis-transformation} to the the level of defect insertions,
	\begin{align}
		\left\langle \prod_{j=1}^{n}
		U_{\gamma_j}[C_j]\right\rangle_{\mathcal R_n}
		&=
		\left\langle \prod_{j=1}^{n}
		U_{\gamma_j}[a_j]U_{-\gamma_j}[a_{j-1}]
		\right\rangle_{\mathcal R_n}
		=
		\left\langle \prod_{j=1}^{n-1}
		U_{\gamma_j-\gamma_{j+1}}[a_j]
		\right\rangle_{\mathcal R_n} \nonumber\\
		&=
		\left\langle \prod_{j=1}^{n-1}
		U_{\alpha_j}[a_j]
		\right\rangle_{\mathcal R_n}.
	\end{align}
	Hence, the relative phases $\alpha_i$ appearing in \eqref{charged-moment_n-1} are precisely the $n-1$ independent holonomies along the dual b-cycles.

	In the derivation above we have assumed that we may replace $U_{\gamma_{j}}\left[C_{j}^{+}\right]$ with the $e^{i\gamma_{j}Q_{1}}$ inserted throughout the charge moments \eqref{charged-moment_def2}. However, a significant caveat is that topology dictates that $U_{\gamma_j}[C_j^+]$ must have support outside the interval $A_1$, at least through an infinitesimally small tail. This is the issue of the so-called "sharp-twists" \cite{Casini2020,Benedetti2022}, and it directly relates to the non-factorizable nature of continuum QFT. Namely, the regularization of the endpoints of the symmetry operator acting on the region $A_1$ can lead to UV divergences. The relevant quantities for us are relative entropies or mutual informations, which are explicitly UV-finite in free theories. The results we find below are UV-finite as well and depends only on physical quantities such as the cross-ratio, and are therefore insensitive to such details.\footnote{A complementary geometric observation is that the two defect segments combine into a closed loop, $U_{\gamma_j}[C_j]=U_{\gamma_j}[C_j^+]U_{\gamma_j}^{\dagger}[C_j^-]$. Thus, the complete defect has no free endpoints.}
	
	\subsection{Complementary Charge Measurement and Holevo Information}
	\label{subsection-complemntary-charge}
	It is natural to ask what physical quantity arises when the symmetry defects are placed on the closed $b$-cycles instead. With the subtleties mentioned above, the corresponding partition function can be associated with,
	\begin{align}\label{b-cycle-sym-defect}
		Z_n(0,\beta_i)
		=\Tr_A\Big[&
		\Tr_{\bar A}\big(|\Omega\rangle\langle\Omega|e^{-iQ_{\bar A_1}\beta_1}\big)
		\Tr_{\bar A}\big(e^{iQ_{\bar A_1}\beta_1}|\Omega\rangle\langle\Omega|e^{-iQ_{\bar A_1}\beta_2}\big)
		\cdots\notag\\
		&\cdots
		\Tr_{\bar A}\big(e^{iQ_{\bar A_1}\beta_{n-1}}|\Omega\rangle\langle\Omega|\big)
		\Big].
	\end{align}
	We use a double argument notation to distinguish between the two choices: $Z_n(0,\beta_i)$ for \eqref{b-cycle-sym-defect} above, and $Z_n(\alpha_i,0)$ for the charged moments \eqref{charged-moment_n-1} previously discussed. Recalling Fig.~\ref{fig1}, $\bar A_1=[v_1,u_2]$ is a complementary interval bridging between $A_1$ and $A_2$. With the additional conventions of $Q_{\bar A_1}$ and its charges $\bar q_1$, we introduce the normalized conditional, or post-measurement states,
	\begin{equation}
		\rho_{A|\bar q_1}
		=\frac{1}{p_{\bar q_1}}
		\Tr_{\bar A}\big(|\Omega\rangle\langle\Omega|\Pi_{\bar q_1}\big),
		\qquad
		p_{\bar q_1}=\langle\Omega|\Pi_{\bar q_1}|\Omega\rangle.
	\end{equation}
	We expand the symmetry operators inside \eqref{b-cycle-sym-defect} in charge projectors \eqref{projectors} to find,
	\begin{equation}\label{b-cycle-sym-defect-avg}
		\int\frac{d^{n-1}\beta}{\left(2\pi\right)^{n-1}}
		Z_n(0,\beta_i)
		=\sum_{\bar q_1}p_{\bar q_1}^{\,n}
		\Tr_A\big(\rho_{A|\bar q_1}^{\,n}\big).
	\end{equation}
	The states $\rho_{A|\bar q_1}$ encode the information available in region $A$ after measuring $Q_{\bar A_1}$ with outcome $\bar q_1$. The averaged entanglement entropy following such charge measurements has recently been studied using the replica trick in the context of measurement-induced entanglement \cite{Milekhin2025observableprojected}. 
	The integrated expression \eqref{b-cycle-sym-defect-avg} is in fact equal to the corresponding charged moment of the complementary region. To see this, consider
	\begin{equation}
		\tilde{\rho}_{\bar{A}}
		=\sum_{\bar q_1}\Pi_{\bar q_1}\rho_{\bar A}\Pi_{\bar q_1}
		=\sum_{\bar q_1}p_{\bar q_1}\rho_{\bar A}^{(\bar q_1)},
		\qquad
		\rho_{\bar A}^{(\bar q_1)}
		=\frac{\Pi_{\bar q_1}\rho_{\bar A}\Pi_{\bar q_1}}{p_{\bar q_1}}.
	\end{equation}
	Since the post-measurement global state $\Pi_{\bar q_1}|\Omega\rangle$ is pure, its reductions to $A$ and $\bar A$ have identical nonzero spectra,
	\begin{equation} \label{spectral-equivalence}
		\Tr_A\big(\rho_{A|\bar q_1}^{\,n}\big)
		=\Tr_{\bar A}\big((\rho_{\bar A}^{(\bar q_1)})^n\big).
	\end{equation}
	Consequently,
	\begin{equation} \label{complementary-duality}
		\int\frac{d^{n-1}\beta}{\left(2\pi\right)^{n-1}}
		Z_n(0,\beta_i)
		=\Tr_{\bar A}\big(\tilde{\rho}_{\bar{A}}^n\big).
	\end{equation}
	
	This also provides an alternative operational interpretation of the EOP in terms of Holevo information; In contrast to the orthogonal charge sectors $\rho_{A}^{\left(q_{1}\right)}$ obtained by measuring $Q_1$ inside $A$, the conditional states $\rho_{A|\bar q_1}$ resulting from a measurement outside $A$ need not be orthogonal. Consequently, the measurement outcome cannot generally be determined with certainty from observables in $A$. This is quantified by the Holevo information,
	\begin{equation}\label{holevodef}
		\chi(Q_{\bar{A}_1}:A)
		=S(\rho_A)-\sum_{\bar q_1}p_{\bar q_1}S(\rho_{A|\bar q_1}).
	\end{equation}
	Using the orthogonality of the charge sectors $\rho_{\bar{A}}^{\left(\bar{q}_{1}\right)}$ and the spectral equivalence \eqref{spectral-equivalence}, we may then write the EOP as,
	\begin{equation}
		S\big(\rho_{\bar{A}}\,\big\|\,\tilde{\rho}_{\bar{A}}\big)=S\big(\tilde{\rho}_{\bar{A}}\big)-S\left(\rho_{\bar{A}}\right)=S_{\mathrm{cl}}-\chi\left(Q_{\bar{A}_{1}}:A\right).
	\end{equation}
	Where $S_{\mathrm{cl}}=-\sum_{\bar q_1}p_{\bar q_1}\log p_{\bar q_1}$ is the classical entropy of the measured charge distribution, which is an upper bound for the Holevo information.
	
	We may now ask what happens when symmetry defects are inserted along both families of canonical cycles. Since in such case the symmetry defects intersect, we need to take into account possible anomalous phases. We leave this case for future study and restrict to non-anomalous symmetries, for which the ordering of the defects is unambiguous. For example, while a single chiral fermion suffices for the cases above, now we should consider a Dirac fermion with both chiralities and the anomaly-free vector $U\!\left(1\right)$ symmetry (or equivalently the axial one). Combining the previous constructions, we deduce that this should correspond to,
	\begin{align}\label{double-cycle-sym-defect}
		Z_n(\alpha_i,\beta_i)
		=\Tr_A\Big[&
		e^{-i\gamma_1 Q_1}
		\Tr_{\bar A}\big(|\Omega\rangle\langle\Omega|e^{-i\beta_1 Q_{\bar A_1}}\big)
		e^{i\gamma_1 Q_1}
		\notag\\
		&\quad\times e^{-i\gamma_2 Q_1}
		\Tr_{\bar A}\big(e^{i\beta_1 Q_{\bar A_1}}|\Omega\rangle\langle\Omega|e^{-i\beta_2 Q_{\bar A_1}}\big)
		e^{i\gamma_2 Q_1}\cdots\notag\\
		&\quad\cdots\times e^{-i\gamma_n Q_1}
		\Tr_{\bar A}\big(e^{i\beta_{n-1} Q_{\bar A_1}}|\Omega\rangle\langle\Omega|\big)
		e^{i\gamma_n Q_1}\Big].
	\end{align}
	Here the $\gamma_j$ implement the charge projection in $A_1$ on each replica sheet, while the $\beta_i$ implement the complementary charge projection in $\bar A_1$. This leads to (recall ${\alpha_i=\gamma_i-\gamma_{i+1}}$),
	\begin{align}\label{double-cycle-sym-defect-avg}
		\int\frac{d^{n-1}\alpha}{\left(2\pi\right)^{n-1}}\frac{d^{n-1}\beta}{\left(2\pi\right)^{n-1}}Z_{n}\left(\alpha_{i},\beta_{i}\right)=\sum_{\bar{q}_{1}}p_{\bar{q}_{1}}^{\,n}\Tr_{A}\left(\tilde{\rho}_{A|\bar{q}_{1}}^{n}\right).
	\end{align}
	With $\tilde{\rho}_{A|\bar{q}_{1}}=\int_{-\pi}^{\pi}\frac{d\gamma}{2\pi}e^{i\gamma Q_{1}}\rho_{A|\bar{q}_{1}}e^{-i\gamma Q_{1}}$. This allows us to quantify how much information about the outcome of the complementary charge measurement $Q_{\bar A_1}$ is retained when measurements within $A$ are restricted to the additive neutral algebra $\mathcal{N}(A_1)\otimes\mathcal{N}(A_2)$, and are therefore insensitive to the charge-transfer correlations. For that aim we define the following Holevo information,
	\begin{equation}
		\chi^{\mathcal{E}}\left(Q_{\bar{A}_{1}}:A\right)=S\left(\tilde{\rho}_{A}\right)-\sum_{\bar{q}_{1}}p_{\bar{q}_{1}}S\left(\tilde{\rho}_{A|\bar{q}_{1}}\right).
	\end{equation}
	The loss relative to the full algebra is
	\begin{equation} \label{Holevo-diff}
		\chi\left(Q_{\bar{A}_{1}}:A\right)
		-\chi^{\mathcal{E}}\left(Q_{\bar{A}_{1}}:A\right)
		=\sum_{\bar{q}_{1}}p_{\bar{q}_{1}}S\left(\rho_{A|\bar{q}_{1}}\,\big\|\,\tilde{\rho}_{A|\bar{q}_{1}}\right)-S\left(\rho_{A}\,\big\|\,\tilde{\rho}_{A}\right)\geq0.
	\end{equation}
	With the positivity following from joint convexity of relative entropy. Formally, we may retrieve this quantity from,
	\begin{equation} \label{Holevo-diff-replica}
			\chi\left(Q_{\bar{A}_{1}}:A\right)
			-\chi^{\mathcal E}\left(Q_{\bar{A}_{1}}:A\right)
			=\lim_{n\rightarrow1}\frac{1}{1-n}
			\log\!\left[
			\frac{
				Z_{n}\left(0,0\right){\displaystyle \iint Z_{n}\left(\alpha_{i},\beta_{i}\right)}
			}{
				\int Z_{n}\left(\alpha_{i},0\right)\int Z_{n}\left(0,\beta_{i}\right)
			}
			\right].
	\end{equation}

	\section{Free Fermion \& Compact Boson: Exact Solutions}
	\label{section-exactresults}
	
	 We have established in section \ref{subsection-replica-and-symmdef} that the insertion of the symmetry defects in the charged moments \eqref{charged-moment_def} introduces holonomies, or twisting conditions along the independent cycles of the genus $n-1$ Riemann surface $\mathcal{R}_n$. Such twisted partition functions for free massless fermions and bosons were computed long ago.\footnote{Additionally, for the Ising model treated in this context, see \cite{Benedetti:2026nlt}.} We will mostly use \cite{AlvarezGaume1986}, however there are many other related works on this subject. Fortunately, due to the unique properties of these free theories, the knowledge of the untwisted partition function $Z_n(0,0)$ allows to reconstruct the whole family of $Z_n(\alpha_i,\beta_i)$. The untwisted partition function is simply $\Tr_A(\rho_{A}^{n})$, which is directly related to the mutual information of the two intervals. These are well known for fermions \cite{Casini2009a}, and for the compact boson \cite{Calabrese2009}. The geometric data determining the Riemann surface $\mathcal{R}_n$ is the Riemann period matrix, which for the replica construction is controlled by the conformal cross-ratio. Let $\ell_1=v_1-u_1$ and $\ell_2=v_2-u_2$ be the interval lengths, and let $r=u_2-v_1$ be their separation. The conformal cross-ratio is
	 \begin{equation}\label{cross-ratio}
	 	\eta=\frac{(v_1-u_1)(v_2-u_2)}{(u_2-u_1)(v_2-v_1)}
	 	=\frac{\ell_1\ell_2}{(\ell_1+r)(\ell_2+r)},\qquad 0<\eta<1.
	 \end{equation}
	 To define the period matrix, we recall the the canonical homology basis $\{a_i,b_i\}_{i=1}^{n-1}$ discussed in section \ref{subsection-replica-and-symmdef}. One chooses a basis of holomorphic one-forms $\omega_j$ normalized by,
	 \begin{equation}
	 	\oint_{a_i}\omega_j=\delta_{ij}.
	 \end{equation}
	 The period matrix is then defined by,
	 \begin{equation}
	 	\Gamma_{ij}=\oint_{b_i}\omega_j.
	 	\label{period_matrix_definition}
	 \end{equation}
	 For this replica surface and homology basis, it is known explicitly \cite{Calabrese2009,Calabrese2011,Coser2016},
	 \begin{equation}
	 	(\Gamma_n)_{ij}
	 	=\frac{2i}{n}\sum_{k=1}^{n-1}
	 	\sin\left(\frac{\pi k}{n}\right)
	 	\frac{{}_2F_1\left(\frac{k}{n},1-\frac{k}{n};1;1-\eta\right)}
	 	{{}_2F_1\left(\frac{k}{n},1-\frac{k}{n};1;\eta\right)}
	 	\cos\left[\frac{2\pi k}{n}(i-j)\right].
	 	\label{replica_period_matrix}
	 \end{equation}
	 For $0<\eta<1$, the matrix is purely imaginary, $\Gamma_n=iY_n$, with $Y_n$ real, symmetric, and positive definite. 
	 For a single complex chiral fermion,
	\begin{equation} \label{free_fermion_Z0}
		\Tr_{A}\left(\rho_{A}^{n}\right)=Z_{n}^{\left(F\right)}\left(0,0\right)=c_{n}^{2}\left[\frac{\ell_{1}\ell_{2}}{a^{2}}(1-\eta)\right]^{\frac{1-n^{2}}{12n}},
	\end{equation}
	where $a$ is a UV cutoff and $c_n$ is a non-universal normalization constant which is theory dependent \cite{Calabrese2004}. The dependence on $\ell_1\ell_2$ reproduces the single-interval entropies, while the factor involving $1-\eta$ yields the R\'enyi mutual information,
	\begin{equation} \label{free_fermion_renyi_MI}
		I_{A}^{(n)}=\frac{1}{1-n}\log\frac{\Tr_{A_{1}}\left(\rho_{A_{1}}^{n}\right)\Tr_{A_{2}}\left(\rho_{A_{2}}^{n}\right)}{\Tr_{A}\left(\rho_{A}^{n}\right)}=-\frac{n+1}{12n}\log\left(1-\eta\right).
	\end{equation}
	For a Dirac fermion consisting of two chiral components, this result is multiplied by a factor of two. For the compact boson, the untwisted partition function is instead given by,
	\begin{equation} \label{compact-boson-zero-moment}
		\Tr_{A}\left(\rho_{A}^{n}\right)=Z_{n}^{\left(B\right)}\left(0,0\right)=c_{n}^{2}\left[\frac{\ell_{1}\ell_{2}}{a^{2}}(1-\eta)\right]^{\frac{1-n^{2}}{6n}}\frac{\Theta\left(0|\lambda\Gamma_{n}\right)\Theta\left(0|\frac{\Gamma_{n}}{\lambda}\right)}{\Theta\left(0|\Gamma_{n}\right)^{2}}.
	\end{equation}
	Here $\lambda=2\pi R^2$, where $R$ is the compactification radius.\footnote{In \cite{Calabrese2009}, the conformal cross-ratio is denoted by $x$, while $\eta$ denotes the compactification parameter corresponding to our $\lambda$, up to conventions.} The Riemann theta functions appearing in \eqref{compact-boson-zero-moment} encode the contributions of the nontrivial winding sectors around the cycles of the replica surface. The insertion of symmetry defects modifies these contributions, which are then conveniently described by theta functions with characteristics. The two are defined as follows, where the latter is the theta function with characteristics,
	\begin{equation}
		\Theta(\boldsymbol{z}|\Gamma)=\sum_{\boldsymbol{m}\in\mathbb{Z}^{n-1}}e^{i\pi\boldsymbol{m}^{{\rm T}}\Gamma\boldsymbol{m}+2\pi i\boldsymbol{m}^{{\rm T}}\boldsymbol{z}},
		\label{riemann_theta}
	\end{equation}
	\begin{equation}
		\Theta\!\begin{bmatrix}\boldsymbol{\beta}\\
			\boldsymbol{\alpha}
		\end{bmatrix}(\boldsymbol{z}|\Gamma)=\sum_{\boldsymbol{m}\in\mathbb{Z}^{n-1}}e^{i\pi\left(\boldsymbol{m}+\boldsymbol{\beta}\right)^{{\rm T}}\Gamma\left(\boldsymbol{m}+\boldsymbol{\beta}\right)+2\pi i\left(\boldsymbol{m}+\boldsymbol{\beta}\right)^{{\rm T}}\left(\boldsymbol{z}+\boldsymbol{\alpha}\right)}.
		\label{riemann_theta_characteristic}
	\end{equation}
	
	\subsection{Chiral and Dirac Fermions}
	In addition to any $U\!\left(1\right)$ holonomies, a fermionic theory on a Riemann surface requires a choice of spin structure. This specifies periodic (Ramond) or antiperiodic (Neveu--Schwarz) boundary conditions around the independent cycles, represented by half-integer characteristics with components $0$ or $1/2$. A genus-$g$ surface admits $2^{2g}$ distinct spin structures. The original, unreplicated manifold is the complex plane, which has genus zero and admits a unique spin structure. From that, the replica surface $\mathcal{R}_n$ inherits the spin structure with vanishing characteristic $(0,0)$ \cite{Headrick2013,Coser2016}. A continuous $U\!\left(1\right)$ holonomy is an additional twist relative to the underlying spin structure, which in this case is the zero characteristic. We write $\alpha_i=2\pi\tilde\alpha_i$ and $\beta_i=2\pi\tilde\beta_i$ to adhere with conventional normalization. The twisted fermionic partition function $Z_{n}(\alpha_{i},\beta_i)$ is then given by \cite{AlvarezGaume1986},
	\begin{equation} \label{Quillen}
		Z_{n}(\alpha_{i},\beta_i)=C[\mathcal{R}_{n}]\,\Theta\!\begin{bmatrix}\tilde{\boldsymbol\beta}\\
			\tilde{\boldsymbol\alpha}
		\end{bmatrix}(0|\Gamma_{n})
	\end{equation}
	where $C[\mathcal R_n]$ is independent of the holonomies. By straightforward integration \eqref{theta_alpha_average}, we may then easily obtain the charged moments needed for the EOP \eqref{EOP-from-ratio} from \eqref{riemann_theta_characteristic} and \eqref{Quillen},
	\begin{equation} \label{fermion_projected}
		\frac{\Tr_{A}\left(\tilde{\rho}_{A}^{n}\right)}{\Tr_{A}\left(\rho_{A}^{n}\right)}=\int\frac{d^{n-1}\alpha}{\left(2\pi\right)^{n-1}}\frac{Z_{n}\!\left(\alpha_{i},0\right)}{Z_{n}\!\left(0,0\right)}=\frac{1}{\Theta\left(0|\Gamma_{n}\right)}.
	\end{equation}
	In this particular case, we can proceed further and perform the analytic continuation in $n$, owing to the following identity \cite{Calabrese2009},
	\begin{equation}
		\Theta(0|\Gamma_n)^2
		=
		\prod_{k=1}^{n-1}
		{}_2F_1\left(
		\frac{k}{n},
		1-\frac{k}{n};
		1;\eta
		\right).
		\label{theta_hypergeometric_product}
	\end{equation}
	Which allows to write \eqref{fermion_projected} as,
	\begin{equation}
		I_A^{(n)}-\tilde I_A^{(n)}
		=
		\frac{1}{2(n-1)}
		\sum_{k=1}^{n-1}
		\log
		{}_2F_1\left(
		\frac{k}{n},
		1-\frac{k}{n};
		1;\eta
		\right).
		\label{renyi_eop_fermion}
	\end{equation}
	The finite sum in \eqref{renyi_eop_fermion} can be analytically continued in $n$ by rewriting it as a contour integral using the Sommerfeld--Watson method,
	\begin{equation} \label{fermion_SW_EOP}
		I_A^{(n)}-\tilde I_A^{(n)}
		=-\frac{in}{2(n-1)}\int_0^\infty ds\,
		\left[\coth(\pi n s)-\coth(\pi s)\right]
		\log\!\left[
		\frac{{}_2F_1(1+is,-is;1;\eta)}
		{{}_2F_1(1-is,is;1;\eta)}\right].
	\end{equation}
	This agrees precisely with the result of \cite{PhysRevD.98.125008}, where the Rényi mutual information was computed by a different method using the neutral chiral current algebra. In their notation, our result is the negative of the correction $U_n(\eta)$.

	We proceed to show that the averaging over the phases $\beta_i$ associated with the $b$-cycles gives the charged moments for the complementary region, as was formally derived in \eqref{complementary-duality}. Using \eqref{theta_b_gaussian} gives,
	\begin{equation}\label{fermion_complement_projected}
		\int\frac{d^{n-1}\beta}{(2\pi)^{n-1}}\frac{Z_{n}\!\left(0,\beta_{i}\right)}{Z_{n}\!\left(0,0\right)}=\frac{1}{\Theta\left(0|-\Gamma_{n}^{-1}\right)}.
	\end{equation}
	The matrix $-\Gamma_{n}^{-1}(\eta)$ is related to $\Gamma_{n}(1-\eta)$ by a change in the homology basis which leaves the theta functions invariant: ${\Theta(0|-\Gamma_n^{-1})=\Theta(0|\Gamma_n(1-\eta))}$. This fact can be directly verified from \eqref{replica_period_matrix}, and it geometrically corresponds to a modular transformation that exchanges the roles of the $a$- and $b$-cycles, as well as $\eta\leftrightarrow 1-\eta$. Comparing with \eqref{fermion_projected}, we indeed recover the charged moment of the complementary region.

	Finally, averaging over the holonomies associated with both families of canonical cycles allows us to evaluate the Holevo-information difference introduced in Section~\ref{subsection-replica-and-symmdef}. As discussed above, to avoid possible anomalous phases we now consider a non-chiral Dirac fermion with the vector $U\!\left(1\right)$ symmetry, rather than a single chiral fermion. The two chiralities contribute complex-conjugate factors, giving the partition function ratio,
	\begin{equation}\label{fermion_dirac_both_cycles}
		\frac{Z_n^{(D)}(\alpha_i,\beta_i)}{Z_n^{(D)}(0,0)}
		=
		\frac{
			\left|\Theta\!\begin{bmatrix}
				\tilde{\boldsymbol\beta}\\
				\tilde{\boldsymbol\alpha}
			\end{bmatrix}(0|\Gamma_n)\right|^2
		}{\Theta(0|\Gamma_n)^2}.
	\end{equation}
	Where $Z_n^{(D)}$ refers to Dirac theory. The symmetry averages over the $a$-cycles, the $b$-cycles, and both cycles are respectively given in \eqref{thetasqr_alpha_average},\eqref{thetasqr_b_gaussian},\eqref{thetasqr_ab_gaussian}. Using these formulas we find,
	\begin{align}\label{fermion_dirac_integrated}
		\int\frac{d^{n-1}\alpha}{(2\pi)^{n-1}}\frac{Z_{n}^{(D)}(\alpha_{i},0)}{Z_{n}^{(D)}(0,0)}
		&=\frac{\Theta(0|2\Gamma_n)}{\Theta(0|\Gamma_n)^2},
		\notag\\
		\int\frac{d^{n-1}\beta}{(2\pi)^{n-1}}\frac{Z_{n}^{(D)}(0,\beta_{i})}{Z_{n}^{(D)}(0,0)}
		&=\frac{\Theta(0|\Gamma_n/2)}
		{\sqrt{\det(-2i\Gamma_n)}\,\Theta(0|\Gamma_n)^2},
		\notag\\
		\int\frac{d^{n-1}\alpha}{(2\pi)^{n-1}}\frac{d^{n-1}\beta}{(2\pi)^{n-1}}\frac{Z_{n}^{(D)}(\alpha_{i},\beta_{i})}{Z_{n}^{(D)}(0,0)}
		&=\frac{1}
		{\sqrt{\det(-2i\Gamma_n)}\,\Theta(0|\Gamma_n)^2}.
	\end{align}
	Substituting these into \eqref{Holevo-diff-replica}, we obtain,
	\begin{equation}\label{fermion_holevo_theta}
		\chi(Q_{\bar A_1}:A)
		-\chi^{\mathcal E}(Q_{\bar A_1}:A)
		=
		\lim_{n\to1}\frac{1}{n-1}
		\log\left[
		\frac{\Theta(0|2\Gamma_n)\Theta(0|\frac{\Gamma_{n}}{2})}
		{\Theta(0|\Gamma_n)^2}
		\right].
	\end{equation}
	Interestingly, the theta-function ratio appearing in this result has a familiar interpretation in the context of bosonization. We notice that the same expression appears in \eqref{compact-boson-zero-moment}, with $\lambda=2$ or equivalently $\lambda=1/2$ (which are related by T-duality). In particular, by combining \eqref{compact-boson-zero-moment} and \eqref{free_fermion_Z0}, the difference between the R\'enyi mutual informations of the compact boson and the Dirac fermion with a fixed spin structure is given by,
	\begin{equation}\label{boson_fermion_renyi_difference}
		I_{{\rm boson}}^{(n)}(\eta;\lambda=2)-I_{{\rm Dirac}}^{(n),\,{\rm fixed- spin}}(\eta)=\frac{1}{n-1}\log\!\left[\frac{\Theta(0|2\Gamma_{n})\Theta(0|\frac{\Gamma_{n}}{2})}{\Theta(0|\Gamma_{n})^{2}}\right].
	\end{equation}
	The values $\lambda=1/2,2$ correspond to the compactification radii associated with bosonization of the Dirac fermion. Importantly, the bosonic theory is dual to the Dirac theory with gauged fermion parity $\mathbb Z_2^F$, whose partition function \eqref{compact-boson-zero-moment} involves an appropriate sum over spin structures on the replica surface, as can be further seen from theta function identities \cite{AlvarezGaume1986,SciPostPhys.7.1.007,Headrick2013,Mukhi2018,Gaur2024a}. This differs from the Dirac fermion with the unique fixed spin-structure that gives \eqref{free_fermion_Z0}, which uses only the vanishing characteristic. Consequently, the Holevo-information difference can be expressed as,
	\begin{equation}\label{fermion_holevo_MI_difference}
		\chi(Q_{\bar{A}_{1}}:A)-\chi^{\mathcal{E}}(Q_{\bar{A}_{1}}:A) = I_{{\rm Dirac}}^{\mathbb{Z}_{2}^{F}-{\rm gauged}}\left(\eta\right)-I_{{\rm Dirac}}^{{\rm fixed-spin}}\left(\eta\right).
	\end{equation}
	
	The right hand side of \eqref{boson_fermion_renyi_difference} is in fact positive (see  \ref{APP-theta-positivity}) for every integer $n>1$. The non-negativity of the Holevo information difference \eqref{Holevo-diff} implies that \eqref{fermion_holevo_MI_difference} should be positive as well. However, performing the analytical continuation for an arbitrary cross-ratio is a known technical challenge. Hence, we resort to some specific regimes where the analytic continuation is achievable. Using the asymptotic expansions derived in \cite{Calabrese2011},
	\begin{equation} \label{Holevo-diff-Dirac}
		\chi(Q_{\bar{A}_{1}}:A)-\chi^{\mathcal{E}}(Q_{\bar{A}_{1}}:A)=\begin{cases}
			\frac{\pi}{8}\sqrt{\eta}, & \eta\rightarrow0\\
			\frac{\pi}{8}\sqrt{1-\eta}, & \eta\rightarrow1
		\end{cases}
	\end{equation}
	Which is manifestly positive, at near separation limit ($\eta\sim 1$) and the far separation limit ($\eta\sim 0$) of the two intervals. The equality of the two asymptotic limits follows from the symmetry of \eqref{fermion_holevo_theta} under $\eta\leftrightarrow1-\eta$ \cite{Calabrese2011}.
	
	\subsection{Compact Boson}
	\label{sec:compact_boson}
	The free massless compact boson, $\phi\sim\phi+2\pi R$, is given by,
	\begin{equation}
		S[\phi]=\frac{1}{2}\int d\phi\wedge *d\phi.
		\label{compact_boson_action}
	\end{equation}
	The theory possesses the standard shift symmetry which we call $U\!\left(1\right)_T$. We normalize the corresponding symmetry transformation by
	\begin{equation}
		U_{T}\left(\alpha\right):\phi\longrightarrow\phi+R\alpha,\qquad\alpha\sim\alpha+2\pi.
	\end{equation} 
	At the fermionic compactification radius $R=\frac{1}{\sqrt{4\pi}}$, the corresponding Noether current $j_\mu^T=R\partial_\mu\phi$ is related to the chiral currents through the decomposition $\phi=\varphi(z)+\bar\varphi(\bar z)$,
	\begin{equation}
		J(z)=\frac{i}{\sqrt{\pi}}\partial\varphi,
		\qquad
		\bar J(\bar z)=-\frac{i}{\sqrt{\pi}}\bar\partial\bar\varphi,
		\label{BosonCurrents}
	\end{equation}
	and one may locally bosonize the Dirac fermion according to
	\begin{equation} \label{bosonoization}
		\psi(z)=\frac{1}{\sqrt{2\pi}}e^{-i\sqrt{4\pi}\varphi(z)},
		\qquad
		\psi^\dagger(z)=\frac{1}{\sqrt{2\pi}}e^{i\sqrt{4\pi}\varphi(z)}.
	\end{equation}
	Additionally, we denote by $U\!\left(1\right)_W$ the topological winding symmetry generated by the current $j_W^\mu=\frac{1}{2\pi R}\epsilon^{\mu\nu}\partial_\nu\phi$. It is convenient to introduce the dual boson,
	\begin{equation}
		\tilde{\phi}=\varphi(z)-\bar\varphi(\bar z).
	\end{equation}
	The dual boson has periodicity $\tilde{\phi}\sim\tilde{\phi}+\frac{1}{R}$, and it is shifted by the winding symmetry: $U_{W}\left(\alpha\right):\tilde{\phi}\mapsto\tilde{\phi}+\alpha/(2\pi R)$. The two symmetries are exchanged by T-duality, $R\rightarrow\frac{1}{2\pi R}$. The operators charged under $U\!\left(1\right)_T$ and $U\!\left(1\right)_W$ are the vertex operators,
	\begin{equation}\label{compact_vertex_operators}
		V_{m,w}(z,\bar{z})=:e^{i\frac{m}{R}\phi+i2\pi Rw\tilde{\phi}}:
		\qquad m,w\in\mathbb{Z}.
	\end{equation}
	We define $\lambda=2\pi R^2$ (so that T-duality acts as $\lambda\leftrightarrow\lambda^{-1}$). 
	The vertex operators are charged under the two symmetries,
	\begin{equation}\label{boson_vertex_symmetries}
		U_T(\alpha):\ V_{m,w}\mapsto e^{im\alpha}V_{m,w},
		\qquad
		U_W(\alpha):\ V_{m,w}\mapsto e^{iw\alpha}V_{m,w}.
	\end{equation}
	In particular, $V_{0,w}$ are neutral under the shift symmetry, whereas $V_{m,0}$ are neutral under the winding symmetry. The currents and their descendants are neutral under both symmetries. Thus, the two symmetries define distinct neutral subalgebras, and we may consider the corresponding projections of the reduced state. As in Section~\ref{section-EOP}, these projections are implemented by averaging over symmetry transformations localized in the first interval $A_1$ with the appropriate charges,
	\begin{equation}\label{boson_two_expectations}
		\tilde{\rho}_{A}^{T}
		=\int_{-\pi}^{\pi}\frac{d\alpha}{2\pi}e^{-i\alpha Q_{1}^{(T)}}\rho_{A}e^{i\alpha Q_{1}^{(T)}},
		\qquad
		\tilde{\rho}_{A}^{W}
		=\int_{-\pi}^{\pi}\frac{d\alpha}{2\pi}e^{-i\alpha Q_{1}^{(W)}}\rho_{A}e^{i\alpha Q_{1}^{(W)}}.
	\end{equation}
	As before, their EOPs are also given by the corresponding differences of mutual informations,
	\begin{equation}
			S\left(\rho_{A}\|\tilde{\rho}_{A}^{T}\right)
			=I_A-\tilde I_{A,\rm T},
			\qquad
			S\left(\rho_{A}\|\tilde{\rho}_{A}^{W}\right)
			=I_A-\tilde I_{A,W}.
	\end{equation}

	Similarly to the fermionic case \eqref{Quillen}, the twisted bosonic partition function dependence on the holonomies is encoded in theta functions \cite{AlvarezGaume1986,Dixon1987,Calabrese2009}. For the boson, this structure arises from summing over classical solutions with different topological winding numbers around the canonical cycles of $\mathcal{R}_n$,
	\begin{equation}\label{boson_classical_quantum}
		Z_n=Z_{{\rm qu}}\,Z_{{\rm cl}},
		\qquad
		Z_{{\rm cl}}=\sum_{\boldsymbol m,\boldsymbol n\in\mathbb Z^{n-1}}
		e^{-S_{\rm cl}(\boldsymbol m,\boldsymbol n)}.
	\end{equation}
	This results from expanding, $\phi=\phi_{\rm cl}+\phi_{\rm qu}$, where the classical field carries winding numbers $\oint_{a_i}d\phi_{\rm cl}=2\pi R m_i$ and $\oint_{b_i}d\phi_{\rm cl}=2\pi R n_i$, whereas the quantum fluctuations $\phi_{\rm qu}$ have none. Crucially, the quantum fluctuation determinant $Z_{qu}$ is independent of the holonomies thanks to the simple structure of the action. We first consider the shift symmetry $U\!\left(1\right)_T$. Inserting the symmetry along the $a_i$-cycle modifies the twisting condition across the dual $b_i$-cycle by $\oint_{b_i}d\phi_{\rm cl}=2\pi Rn_i+R\alpha_i$. After direct evaluation of the classical action, which is reviewed in \ref{APP-boson-sectors}, one obtains,
	\begin{equation}\label{boson_classical_lattice_factors}
		Z_{n}^{(T)}(\alpha_{i},0)=Z_{qu}\;\Theta(0|\lambda\Gamma_{n})\,\Theta\!\begin{bmatrix}\boldsymbol{0}\\
			\tilde{\boldsymbol{\alpha}}
		\end{bmatrix}\left(0|\Gamma_{n}/\lambda\right).
	\end{equation}
	For our purposes the quantum term is not required, since it cancels in the ratio \eqref{EOP-from-ratio}. Using \eqref{theta_alpha_average}, this simply gives,
	\begin{equation}\label{boson_translation_projected}
		\frac{\Tr_{A}\left[(\tilde{\rho}_{A}^{T})^{n}\right]}{\Tr_{A}(\rho_{A}^{n})}=\int\frac{d^{n-1}\alpha}{(2\pi)^{n-1}}\frac{Z_{n}^{(T)}\left(\alpha_{i},0\right)}{Z_{n}^{(T)}\left(0,0\right)}=\frac{1}{\Theta\left(0|\Gamma_{n}/\lambda\right)}.
	\end{equation}
	Consequently,
	\begin{equation}\label{translation_EOP}
		I^{(n)}-\tilde{I}_{{\rm T}}^{(n)}=\frac{1}{n-1}\log\Theta\left(0|\Gamma_{n}/\lambda\right).
	\end{equation}
	At the self-dual radius $\lambda=1$, this result coincides with the fermionic result \eqref{fermion_projected}, for which analytic continuation in $n$ is possible thanks to identity \eqref{theta_hypergeometric_product}. The corresponding result for the winding symmetry $U\!\left(1\right)_W$ is similarly obtained from T-duality,
	\begin{equation}\label{winding_EOP}
		I^{(n)}-\tilde I_{W}^{(n)}
		=\frac{1}{n-1}\log\Theta(0|\lambda\Gamma_n).
	\end{equation}
	We note that for both symmetries, in the limit of nearly touching intervals $\eta\approx 1$, the universal behavior anticipated in \cite{Casini2020} is displayed for every integer $n>1$,
	\begin{equation}
		\tilde{I}_{A}^{\left(n\right)}=I_{A}^{\left(n\right)}-\frac{1}{2}\log\left(-\log\left(1-\eta\right)\right)+\dots
	\end{equation}
	We shall study this limit further through more general methods in section \ref{section-nonAbelian}.
	
	We can see that the correspondence between the charged moments of the complementary region and the averaged partition function over the b-cycles \eqref{complementary-duality} holds in this case as well. Using \eqref{theta_b_gaussian},
	\begin{equation}
		\begin{aligned}
			&\int\frac{d^{n-1}\beta}{(2\pi)^{n-1}}
			\frac{Z^{(T)}_n(0,\beta_i)}
			{Z^{(T)}_n(0,0)}
			=
			\frac{1}{
				\sqrt{\det(-i\lambda\Gamma_n(\eta))}
				\,\Theta(0|\lambda\Gamma_n(\eta))
			} \\
			&=
			\frac{1}{\Theta(0|-\Gamma_n(\eta)^{-1}/\lambda)}
			=
			\frac{1}{\Theta(0|\Gamma_n(1-\eta)/\lambda)}
			=
			\frac{\Tr_{\bar A}[(\tilde\rho_{\bar A}^{T})^n]}
			{\Tr_{\bar A}(\rho_{\bar A}^n)}.
		\end{aligned}
	\end{equation}

	Finally, we consider the Holevo-information difference \eqref{Holevo-diff} for this case. For the compact boson, occurs a factorization,
	\begin{equation}\label{boson_Holevo_factorization}
		\frac{Z_{n}(\alpha_{i},\beta_{i})}{Z_{n}(0,0)}=\frac{Z_{n}(\alpha_{i},0)}{Z_{n}(0,0)}\frac{Z_{n}(0,\beta_{i})}{Z_{n}(0,0)}.
	\end{equation}
	Thus, $\log\frac{Z_{n}(\alpha_{i},\beta_{i})Z_{n}(0,0)}{Z_{n}(\alpha_{i},0)Z_{n}(0,\beta_{i})}=0$ for integer $n>1$, and we can infer that the Holevo difference \eqref{Holevo-diff-replica} vanish as well.\footnote{Although the Holevo difference involves a ratio of moments, introducing an auxiliary Hilbert space allows us to express its numerator and denominator as moments of two density matrices. Their equality at every integer Rényi index implies equality of their spectra, and hence of their von Neumann entropies \cite{DHoker2021,Solodukhin:2011gn}.}
	This is in sharp contrast with the Dirac fermion with fixed spin-structure \eqref{Holevo-diff-Dirac}.

	\subsection{$\mathbb{Z}_N$ Discrete Symmetry}
	\label{sec:Zn_projection}
	
	A significant advantage of this framework is that it also allows us to easily obtain the EOP for discrete symmetries. One of the characterizing features of EOPs of discrete symmetries that sets them apart from continuous ones, is their distinct universal behavior in the limit of nearly touching intervals. While the latter diverges, discrete symmetries approach in the limit a constant, $\log|G|$, where $|G|$ is the order of the group. In more general cases pertaining to generalized symmetries, $|G|$ is replaced by the Jones index \cite{Casini2020,PhysRevD.110.125004,Benedetti:2026nlt}. We therefore consider neutrality under a discrete subgroup $\mathbb Z_N\subset U\!\left(1\right)$, for which the projection is implemented by,
	\begin{equation}\label{ZN_channel}
		\tilde{\rho}_{A}^{\mathbb{Z}_{N}}=\frac{1}{N}\sum_{s=0}^{N-1}e^{-i\frac{2\pi s}{N}Q_{1}}\rho_{A}e^{i\frac{2\pi s}{N}Q_{1}}.
	\end{equation}
	The corresponding neutral algebra contains the $U\!\left(1\right)$-neutral algebra, $\mathcal N_{U\!\left(1\right)}\subset\mathcal N_{\mathbb Z_N}$. It is larger than $\mathcal N_{U\!\left(1\right)}$ since it also admits operators carrying nonzero charges that are multiples of $N$. More generally, $\mathcal N_{\mathbb Z_N}\subset\mathcal N_{\mathbb Z_M}$ whenever $M$ divides $N$. This inclusion of algebras gives us the hierarchy,
	\begin{equation} \label{EOP-hierarchy}
		S\left(\rho_{A}||\tilde{\rho}_{A}^{\mathbb{Z}_{M}}\right)\leq S\left(\rho_{A}||\tilde{\rho}_{A}^{\mathbb{Z}_{N}}\right)\leq S\left(\rho_{A}||\tilde{\rho}_{A}^{U\!\left(1\right)}\right),\quad M\,|\,N.
	\end{equation}
	The twisted partition functions remain unchanged, and only the group average is replaced by a finite sum.  After using  \eqref{theta_ZN_a_average}, the projected moments for the field-shifting symmetry become
	\begin{equation} \label{Z_NT}
		\frac{\Tr_{A}\!\left[\left(\tilde{\rho}_{A}^{\mathbb{Z}_{N,T}}\right)^{n}\right]}{\Tr_{A}\rho_{A}^{n}}=\frac{1}{N^{n-1}}\overset{}{\underset{\boldsymbol{s}\in\mathbb{Z}_{N}^{n-1}}{\sum}}\frac{Z_{n}\left(\frac{2\pi s_{i}}{N},0\right)}{Z_{n}\left(0,0\right)}=\frac{\Theta(0|N^{2}\Gamma_{n}/\lambda)}{\Theta(0|\Gamma_{n}/\lambda)}.
	\end{equation}
	For the winding symmetry we exchange $\lambda\rightarrow \lambda^{-1}$, and for the chiral fermion $\lambda=1$. The period matrix has a purely imaginary part that is real and positive definite, and therefore increasing $N$ suppresses the numerators of \eqref{Z_NT} term by term. This agrees with the hierarchy \eqref{EOP-hierarchy}. Furthermore, the familiar $U\!\left(1\right)$ result \eqref{boson_translation_projected} is recovered as $N\to\infty$.
	
	We can now verify the universal behavior in the nearly touching interval limit mentioned above. As $\eta\to1$, the contributions from all nontrivial group elements vanish, leaving only the identity contribution in the group average. Consequently,
	\begin{equation}
		I_A^{(n)}-\tilde I_{A,\mathbb Z_N}^{(n)}
		\longrightarrow \log N,
	\end{equation}
	Which is indeed the desired limiting behavior with $|\mathbb{Z}_N|=N$.

	The factorization \eqref{boson_Holevo_factorization} means that the Holevo information difference \eqref{Holevo-diff} vanishes for the compact boson also when considering subgroups of $U\!\left(1\right)$. However, for the Dirac fermion the Holevo information difference \eqref{fermion_holevo_MI_difference} is non-vanishing, and as we have seen is closely related to fermion-parity gauging. Interestingly, for even $N$, combining the $\mathbb{Z}_N$ averages \eqref{dirac_ZN_a_even}, \eqref{dirac_ZN_b_even}, and \eqref{dirac_ZN_double_even} give
	\begin{equation} \label{ZN-Holevo-even}
		\frac{Z_{n}\left(0,0\right)\underset{\boldsymbol{s},\boldsymbol{r}\in\mathbb{Z}_{N}^{n-1}}{\sum}Z_{n}\left(\frac{2\pi s_{i}}{N},\frac{2\pi r_{i}}{N}\right)}{\underset{\boldsymbol{s}\in\mathbb{Z}_{N}^{n-1}}{\sum}Z_{n}\left(\frac{2\pi s_{i}}{N},0\right)\underset{\boldsymbol{r}\in\mathbb{Z}_{N}^{n-1}}{\sum}Z_{n}\left(0,\frac{2\pi r_{i}}{N}\right)}=\frac{\Theta(0|\Gamma_{n})^{2}}{\Theta(0|2\Gamma_{n})\Theta(0|\Gamma_{n}/2)},
		\qquad N\ \mathrm{even}.
	\end{equation}
	This expression coincides with the continuous $U\!\left(1\right)$ result \eqref{fermion_holevo_theta} for every even subgroup. For odd $N$, there is no cancellation and the averages instead involve additional sums over half-integer characteristics, and are given in \eqref{dirac_ZN_a_odd}, \eqref{dirac_ZN_b_odd}, and \eqref{dirac_ZN_double_odd}. While the full expression for odd $N$ is rather large, we can study the limits $\eta\rightarrow0,1$. The leading correction to the ratio of averages \eqref{Holevo-diff-replica} for odd $N\geq3$ behaves as
	\begin{equation}
		1-C_{n,N}(1-\eta)^{\frac12+\frac{1}{2N^2}}
		+\dots
	\end{equation}
	By the modular symmetry discussed above, the same behavior holds for $\eta\rightarrow0$, with $1-\eta$ replaced by $\eta$, which can also be verified numerically. The coefficient $C_{n,N}$ can be computed and analytically continued with the methods introduced in \cite{Calabrese2011}, but we don't derive an explicit expression here. The leading behavior is therefore,
	\begin{equation} \label{Holevo-diff-Dirac-Nodd}
		\chi(Q_{\bar A_1}:A)-\chi^{\mathcal E}(Q_{\bar A_1}:A)
		\sim
		\begin{cases}
			c_N\eta^{\frac12+\frac{1}{2N^2}},
			& \eta\rightarrow0,\\
			c_N(1-\eta)^{\frac12+\frac{1}{2N^2}},
			& \eta\rightarrow1.
		\end{cases}
	\end{equation}
	Thus, unlike for even $N$ \eqref{Holevo-diff-Dirac}, the leading power depends explicitly on the order of the symmetry group and approaches the $U\!\left(1\right)$ scaling $1/2$ as $N\rightarrow\infty$.

	To conclude, let us come back to the observation that even $N$ provides the same change in Holevo information as the full $U\!\left(1\right)$ group. For Dirac fermion, fermion parity is generated by $(-1)^F=e^{i\pi Q}$, corresponding to the $\mathbb Z_2$ subgroup of $U\!\left(1\right)$ \cite{SciPostPhys.7.1.007}. This subgroup is contained in every even $\mathbb Z_N$, but in no odd one. Our result therefore suggests that the Holevo information difference is sensitive to fermion-parity, but insensitive to additional projections to smaller neutral subalgebras corresponding to larger even $\mathbb Z_N$ subgroups.

	\section{Non-Abelian Symmetries}
	\label{section-nonAbelian}
	The method of the previous section, where we explored the $U\!\left(1\right)$ symmetry and its subgroups, rests on the simple structure of the partition functions \eqref{Quillen}, \eqref{boson_classical_quantum}. When the theory comes with a non-Abelian symmetry, this structure generally becomes more involved \cite{Bernard1988,Naculich1990,Gawedzki1995,Gaberdiel2009,Robbins2020}. Nevertheless, there are several important regimes where the replica partition function, and consequently the EOP and the Holevo information, simplify and exhibit universal behavior. In this section we study the nearly touching limit $\eta\rightarrow 1$ of the two intervals, and the regime of a large number of fermions $N_F\gg 1$ at a wide range of cross-ratios.\footnote{The same computations can also be applied to free massless bosons, yet we leave this outside the scope of this paper.}

	The charged moments are constructed as previously, where now the Haar average over a discrete, or a Lie group $G$ gives
	\begin{equation} \label{charged-moment-NA-n}
		\Tr_A\left(\tilde{\rho}_{A}^{n}\right)
		=
		\int\frac{d^{n}g}{|G|^{n}}
		\Tr_A\left(
		\rho_A U_{A_1}(g_1^{-1}g_2)
		\rho_A U_{A_1}(g_2^{-1}g_3)
		\dots
		\rho_A U_{A_1}(g_n^{-1}g_1)
		\right).
	\end{equation}
	Above $|G|$ is the volume of the Lie manifold, or the order of $G$ for a finite group. Defining $h_i=g_i^{-1}g_{i+1}$ for $i=1,\ldots,n-1$, and taking $g_n$ as the remaining independent variable, the latter can be integrated out by the invariance of the Haar measure. We obtain
	\begin{equation}
		\label{charged-moments-NA-n-1}
		\Tr_A\left(\tilde{\rho}_{A}^{n}\right)
		=
		\int\frac{d^{n-1}h}{|G|^{n-1}}Z_n(h_i,0),
		\quad
		Z_n(h_i,0)
		=
		\Tr_A\left(
		\rho_A U_{A_1}(h_1)
		\rho_A U_{A_1}(h_2)
		\dots
		\rho_A U_{A_1}(h_n)
		\right),
	\end{equation}
	Where we defined $h_{n}=g_{n}^{-1}g_{1}=\left(\Pi_{i=1}^{n-1}h_{i}\right)^{-1}$.

	Here we find it more convenient to take a standard alternative route \cite{Calabrese2004,Casini2005}, and reformulate the replica problem as a partition function of $n$ decoupled free massless fermions $\psi_{i}$, ($i=1,\dots,n$) living on the original non-replicated manifold, with additional twist operators implementing the appropriate replica monodromy around the entangling points $u_1,v_1,u_2,v_2$ (c.f. Fig.~\ref{fig1}). The replica monodromy cyclically permutes the different copies, $\psi_i\rightarrow\psi_{i-1}$, where for fermions $\psi_1\rightarrow(-1)^{n+1}\psi_n$. Overall, this is implemented by the signed permutation matrix
	\begin{equation} \label{permutation-matrix}
		P=\begin{pmatrix}0 & 1 & 0 & \cdots & 0\\
			0 & 0 & 1 & \cdots & 0\\
			\vdots & 0 & \ddots & \ddots & \vdots\\
			\vdots & 0 & \cdots & 0 & 1\\
			\left(-1\right)^{n+1} & 0 & \cdots & 0 & 0
		\end{pmatrix}.
	\end{equation}
	Assuming the fields transform as a multiplet under a representation $R$ of the group, the symmetry operators in \eqref{charged-moments-NA-n-1} enforce the additional monodromy $\psi_{i,\alpha}\rightarrow R_{\alpha\beta}\left(h\right)\psi_{i,\beta}$. Altogether, there are two different types of monodromies acting in the two intervals,
	\begin{equation}
		\left(M_{1}\right)_{ij,\alpha\beta}=P_{ij}R\left(h_{i}\right)_{\alpha\beta},\qquad\left(M_{2}\right)_{ij,\alpha\beta}=P_{ij}\delta_{\alpha\beta}.
	\end{equation}
	We can then write $Z_{n}\left(h_{i},0\right)$ in terms of the twist operators $\tau_{M}$,
	\begin{equation} \label{4pt-twist}
		Z_{n}\left(h_{i},0\right)=\left\langle \tau_{M_{1}}\left(u_{1}\right)\tau_{M_{1}^{-1}}\left(v_{1}\right)\tau_{M_{2}}\left(u_{2}\right)\tau_{M_{2}^{-1}}\left(v_{2}\right)\right\rangle.
	\end{equation}
	An important fact is that $M_1$ and $M_2$ generally do not commute, also for an Abelian group. More so, since all the group elements in \eqref{charged-moments-NA-n-1} multiply to the identity, then ${M_{2}^{n}=M_{1}^{n}=P^{n}\boldsymbol{1}_{R}=\left(-1\right)^{n+1}\boldsymbol{1}_{R}}$. Hence, the two monodromy matrices share the same eigenvalues: ${\theta_{m}=\frac{2m-n-1}{2n}}$, each with degeneracy $d_R\equiv\text{dim}(R)$.\footnote{See \cite{Benini:2024xjv} for the $U\!\left(1\right)$ case.}

	Crucially, for a general $M=e^{2\pi i\Theta}$, the twist operator behave as a primary field with chiral conformal weight  $h\!\left(M\right)=\frac{1}{2}\Tr\left(\Theta^{2}\right)$. An easy way to see this is to diagonalize $M$ and through bosonization write the twist operator as a product of vertex operators implementing the monodromy $\psi\left(e^{2\pi i}z\right)=e^{2\pi i\theta}\psi$ for each eigenvalue $\theta_m$ of $M$ individually \cite{Casini2005},
	\begin{equation} \label{twist-product}
		\tau_{M}\left(z\right)=:e^{-i\sum_{m=1}^{n}\sum_{\alpha=1}^{d_R}\sqrt{4\pi}\theta_{m}\varphi_{m,\alpha}\left(z\right)}:
	\end{equation}
	It is well known that for \eqref{permutation-matrix}, and by extension for $M_1,M_2$, that 
	\begin{equation} \label{mondromy-conformal-weights}
		{h\!\left(M_1\right)=h\!\left(M_2\right)=\frac{d_{R}}{24}\left(n-\frac{1}{n}\right)}.
	\end{equation}
	This agrees with the standard replica-twist dimension, since $d_R$ is also the central charge of the chiral fermion charged under the representation $R$.
	
	\subsection{Nearly Touching Intervals}
	\label{SEC-small-sep}
	Recall the definitions,
	\begin{equation}
		\ell_{1}=v_{1}-u_{1},\qquad\ell_{2}=v_{2}-u_{2},\qquad r=u_{2}-v_{1}.
	\end{equation}
	\begin{equation}
		\eta=\frac{(v_1-u_1)(v_2-u_2)}{(u_2-u_1)(v_2-v_1)}
		=\frac{\ell_1\ell_2}{(\ell_1+r)(\ell_2+r)},\qquad 0<\eta<1.
	\end{equation}
	Consider the limit of a very near separation of the two intervals: ${r\ll l_{1},l_{2}}$. Or alternatively $\eta\approx1$. For this limit, we use the OPE of the two inner twist operators $\tau_{M_{1}^{-1}}\left(v_{1}\right)\tau_{M_{2}}\left(u_{2}\right)$,
	\begin{equation}
		Z_{n}\left(h_{i},0\right)\approx\frac{\left\langle \tau_{M_{1}}\left(u_{1}\right)\tau_{M_{1}^{-1}M_{2}}\left(v_{1}\right)\tau_{M_{2}^{-1}}\left(v_{2}\right)\right\rangle }{r^{h\left(M_{1}^{-1}\right)+h\left(M_{2}\right)-h\left(M_{1}^{-1}M_{2}\right)}}.
	\end{equation}
	The omitted terms include higher-dimensional operators with the same total monodromy. Using $1-\eta\approx\frac{l_{1}+l_{2}}{l_{1}l_{2}}r$ and the conformal 3-point functions,
	\begin{equation}
		Z_{n}\left(h_{i},0\right)=c\left(h_{i}\right)\frac{l_{1}^{-2h\left(M_{1}\right)}l_{2}^{-2h\left(M_{2}\right)}}{\left(1-\eta\right)^{h\left(M_{1}\right)+h\left(M_{2}\right)-h\left(M_{1}^{-1}M_{2}\right)}}.
	\end{equation}
	Where $c\left(h_{i}\right)$ is an OPE coefficient, which can be rather complicated even for simple monodromy matrices \cite{Lunin2001}. Using the conformal weights \eqref{mondromy-conformal-weights} the expression simplifies,
	\begin{equation} \label{OPE-small}
		Z_{n}\left(h_{i},0\right)=\left(l_{1}l_{2}\left(1-\eta\right)\right)^{\frac{d_{R}}{12}\left(\frac{1}{n}-n\right)}c\left(h_{i}\right)\left(1-\eta\right)^{h\left(M_{1}^{-1}M_{2}\right)}.
	\end{equation}
	We first notice that, $(M_{1}^{-1})_{ij,\alpha\beta}=(P^{-1})_{ij}R(h_{j}^{-1})_{\alpha\beta}$, and consequently we have \linebreak(modulo $n$ in the indices),
	\begin{equation} \label{mondromy-fused}
		\left(M_{1}^{-1}M_{2}\right)_{ij,\alpha\beta}=\delta_{ij}R^{-1}\left(h_{i-1}\right)_{\alpha\beta}.
	\end{equation}
	The first factor in \eqref{OPE-small} simply give rise to the untwisted partition function associated with the regular mutual information. We can therefore write,
	\begin{equation}\label{NA-Haar-OPE}
		\frac{\Tr_{A}(\tilde{\rho}_{A}^{n})}{\Tr_{A}(\rho_{A}^{n})}=\int\frac{d^{n-1}h}{|G|^{n-1}}\frac{Z_{n}(h_{i},0)}{Z_{n}(0,0)}\approx\int\frac{d^{n-1}h}{|G|^{n-1}}c_{n}\!\left(h_{i}\right)e^{h\left(M_{1}^{-1}M_{2}\right)\log\left(1-\eta\right)}.
	\end{equation}
	Since $\log\left(1-\eta\right)$ is negative and arbitrarily large, the integral is dominated by the minima of $h\!\left(M_{1}^{-1}M_{2}\right)$ and can be approximated by a Gaussian integral. The relevant matrix  \eqref{mondromy-fused} is block diagonal in the replica indices, with one block for each group element $h_i$. In particular, its scaling dimension is non-negative and vanishes when all the monodromies $R(h_i)$ are the identity. The dominant saddle is therefore located at $h_i=e$, consistently with the H\"older inequality argument of section \ref{section-EOP}. Thus, it then suffices to evaluate the integrand near $h_i=e$. With that aim, we choose local coordinates $X_{i}^{a}$ adapted to the Lie algebra by choosing Hermitian generators $T_a,\; a=1,\ldots,\text{dim}(G)$. Near the identity, the independent group elements are parameterized by
	\begin{equation}\label{NA-Lie-parameters}
		R\left(h_{i}\right)=e^{iX_{i}^{a}T_{a}},\qquad i=1,\ldots,n-1.
	\end{equation}
	The basis is chosen as standard such that $\Tr_R(T_aT_b)=T(R)\delta_{ab}$, with $T(R)$ the Dynkin index of the representation. In this basis we expand the Haar measure in the same coordinates,
	\begin{equation}
		dh_{i}=\left(1+\mathcal{O}(X_{i}^{2})\right)\,d^{\left(\text{dim}(G)\right)}X_{i},
	\end{equation} We also recall the final group element is not independent, $h_{n}=g_{n}^{-1}g_{1}=\left(\Pi_{i=1}^{n-1}h_{i}\right)^{-1}$. Using the Baker--Campbell--Hausdorff formula,
	\begin{equation}
		R\left(h_{n}\right)=e^{-i\sum_{i=1}^{n-1}X_{i}^{a}T_{a}+\mathcal{O}(X^{2})}.
	\end{equation}
	It follows that
	\begin{align}
		h(M_1^{-1}M_2)
		&=
		\frac{1}{8\pi^2}
		\left[
		\sum_{i=1}^{n-1}
		\Tr_R\left((X_i^aT_a)^2\right)
		+
		\Tr_R\left(
		\left(\sum_{i=1}^{n-1}X_i^aT_a\right)^2
		\right)
		\right]
		+\mathcal O(X^3)
		\nonumber\\
		&=
		\frac{T(R)}{8\pi^2}
		\left[
		\sum_{i=1}^{n-1}\sum_{a=1}^{\dim G}(X_i^a)^2
		+
		\sum_{a=1}^{\dim G}
		\left(\sum_{i=1}^{n-1}X_i^a\right)^2
		\right]
		+\mathcal O(X^3).
	\end{align}
	We may write this more compactly as
	\begin{equation} \label{NA-quadratic-dimension}
		h\left(M_{1}^{-1}M_{2}\right)=\frac{T(R)}{8\pi^{2}}\sum_{i,j=1}^{n-1}\sum_{a=1}^{\text{dim}\left(G\right)}H_{ij}X_{i}^{a}X_{j}^{a}+O(X^{3}),
	\end{equation}
	\begin{equation} \notag
		H_{ij}=\delta_{ij}+1.
	\end{equation}
	Substituting \eqref{NA-quadratic-dimension} into \eqref{NA-Haar-OPE}, we obtain a Gaussian integral over the $(n-1)\dim G$ independent Lie-algebra coordinates. Finally, using additionally $c_n(e)=1$,
	\begin{equation} \label{NA-Gaussian-integral}
		\frac{\Tr_{A}(\tilde{\rho}_{A}^{n})}{\Tr_{A}(\rho_{A}^{n})}\approx\frac{1}{|G|^{n-1}}\int\prod_{i=1}^{n-1}d^{\left(\dim\left(G\right)\right)}X_{i}\;e^{T(R)\log\left(1-\eta\right)H_{ij}X_{i}^{a}X_{j}^{a}/8\pi^{2}}.
	\end{equation}
	The matrix $H$ is positive definite, and the leading behavior is
	\begin{equation}
		\frac{\Tr_A(\tilde{\rho}_A^n)}
		{\Tr_A(\rho_A^n)}
		\propto
		[-\log(1-\eta)]^{-\frac{(n-1)\dim G}{2}}.
	\end{equation}
	We therefore obtain for every $n>1$,\footnote{Some modifications are required if the representation is reducible or non-faithful. We assume that is not the case.}
	\begin{equation}\label{NA-small-separation}
		I_A^{(n)}-\tilde I_A^{(n)}=
		\frac{\dim \left(G\right) }{2}
		\log[-\log(1-\eta)]
		+\mathcal O(1).
	\end{equation}
	Taking the limit $n\rightarrow 1$, the result agrees with the expected universal behavior for a compact connected Lie group \cite{Casini2020,PhysRevD.110.125004}.

	For a finite group $G$, the same argument becomes particularly simple. The Haar integral in \eqref{NA-Haar-OPE} is replaced by a sum over $|G|^{n-1}$ configurations of the group elements $h_i$. The scaling dimension $h(M_1^{-1}M_2)$ vanishes only when $h_i=e$ for all $i=1,\ldots,n$. As $\eta\rightarrow1$, all other configurations are suppressed by positive powers of $1-\eta$, and only the identity configuration survives. Recalling that $c_n(e)=1$, we obtain
	\begin{equation}
		\frac{\Tr_A(\tilde{\rho}_A^n)}{\Tr_A(\rho_A^n)}
		\approx
		\frac{1}{|G|^{n-1}}
		\sum_{h_1,\ldots,h_{n-1}\in G}
		c_n(h_i)(1-\eta)^{h(M_1^{-1}M_2)}
		\approx
		\frac{1}{|G|^{n-1}}.
	\end{equation}
	Consequently giving the universal limit in agreement with \cite{Casini2020,PhysRevD.110.125004},
	\begin{equation}
			I_A^{(n)}-\tilde I_A^{(n)}
			=
			\log|G|
			\qquad \eta\rightarrow1.
	\end{equation}

	We may similarly discuss the EOP for the complementary region $\bar{A}$, which is related by \eqref{complementary-duality} to the group average with the symmetry defects now placed outside the region, in $\bar{A}_1=[v_1,u_2]$ (c.f. Fig.~\ref{fig1}). Since the cross-ratio of $\bar{A}$ is $1-\eta$, the same limit now corresponds to a large interval separation between $\bar{A}_1$ and $\bar{A}_2$. The monodromies now organize differently and the twist fields are schematically specified by,
	\begin{equation}
		Z_{n}\left(0,h_{i}\right)=\left\langle \tau_{P}\left(u_{1}\right)\tau_{P^{-1}R}\left(v_{1}\right)\tau_{PR^{-1}}\left(u_{2}\right)\tau_{P^{-1}}\left(v_{2}\right)\right\rangle .
	\end{equation}
	The OPE of the two inner twists at $v_1$ and $u_2$ now has the identity operator as the leading term, followed by the chiral currents and their descendants. To see the implications, let us first observe the $\eta\approx 1$ expansion for the exact expression that was previously obtained in \eqref{fermion_complement_projected} for a single chiral free fermion charged under $U\!\left(1\right)$,
	\begin{equation}
		\int\frac{d^{n-1}\beta}{(2\pi)^{n-1}}
		\frac{Z^{(F)}_n(0,\beta_i)}{Z^{(F)}_n(0,0)}
		=
		1-\frac{n^2-1}{12n}(1-\eta)
		+\mathcal O((1-\eta)^2).
	\end{equation}
	However, this coincides with the large separation expansion of the regular Rényi mutual information \eqref{free_fermion_renyi_MI}, so we can infer that the group averaging has removed any contributions up to this order. We can therefore extrapolate and write, at least up to this order,
	\begin{equation}\label{NA-b-cycle-average}
		\int\frac{d^{n-1}h}{|G|^{n-1}}
		\frac{Z_n(0,h_i)}{Z_n(0,0)}
		=
		1-\frac{d_R(n^2-1)}{12n}(1-\eta)
		+\mathcal O((1-\eta)^2).
	\end{equation}
	
	Finally, we mention that in this limit the behavior of the Holevo information difference \eqref{Holevo-diff} depends on subleading terms in the OPE and is less universal. We have seen this for example when examining the different asymptotics for the Dirac fermion which depends on whether $N$ is even \eqref{Holevo-diff-Dirac}, or odd \eqref{Holevo-diff-Dirac-Nodd}.
	
	\subsection{Large Number of Fermions}
	Another limit that allows us to probe the entropic order parameter is for a large number of fermions $N_F\gg1$. We consider $N_F$ independent fermion multiplets, each transforming in the representation $R$ of the continuous symmetry group $G$. Naturally, if we take the neutral algebra with respect to the product group  $G^{\otimes N_F}$, and perform the projection as in \eqref{dephasedA}, we will simply have a factorization, $\tilde S_A^{(N_F)}=N_F\tilde S_A$. A more interesting route is to consider the diagonal subgroup $G_D\subset G^{\otimes N_F}$, generated by
	\begin{equation}
		Q_{A_1}^a
		=
		\sum_{f=1}^{N_F}Q_{A_1}^{a,(f)},
		\qquad a=1,\ldots,\text{dim}(G).
	\end{equation}
	Since the vacuum of the free theory is a product state,  $\rho_A=\bigotimes_{f=1}^{N_F}\rho_A^{(f)}$, and once we project the reduced vacuum state to be neutral under the diagonal subgroup $G_D$ we find
	\begin{equation}\label{saddle}
		\Tr\left(\tilde\rho_D^n\right)
		=
		\int\frac{d^{n-1}h}{|G|^{n-1}}
		e^{N_F\log Z_n(h_i,0)},
	\end{equation}
	where $Z_n(h_i,0)$ is the twisted partition function of a single fermion multiplet, defined in \eqref{charged-moments-NA-n-1}. It is therefore natural to perform a saddle expansion utilizing $N_F \gg1$, with the same techniques of section \ref{SEC-small-sep}. The main result is provided in equation \eqref{charged-moment-largeN}.

	Using the Lie-algebra coordinates \eqref{NA-Lie-parameters}, we define the matrix required for this expansion,
	\begin{equation}
		\mathcal K_{ij}^{ab}
		=
		\left.
		-\frac{\partial^2\log Z_n(h_i,0)}
		{\partial X_i^a\partial X_j^b}
		\right|_{X_i^a=0},
		\qquad i,j=1,\ldots,n-1.
	\end{equation}
	As we will see in more detail below, from the symmetry of the vacuum and the normalization $\Tr_R(T_aT_b)=T(R)\delta_{ab}$, the group indices factorize out,
	\begin{equation}
		\mathcal K_{ij}^{ab}
		=
		T(R)\delta_{ab}\mathcal K_{ij}.
	\end{equation}
	A Gaussian expansion around $h_i=e$ then gives,
	\begin{equation}
		\Tr\left(\tilde\rho_D^n\right)
		\approx
		\Tr(\rho_A^n)
		\frac{1}{|G|^{n-1}}
		\left(\frac{2\pi}{N_FT(R)}\right)^{
			\frac{(n-1)\text{dim}(G)}{2}}
		\left(\det\mathcal K\right)^{-\frac{\text{dim}(G)}{2}}.
	\end{equation}
	Consequently,
	\begin{equation}
		\tilde S_A^{(n)}
		=
		N_FS_{A,R}^{(n)}
		+\log|G|
		+\frac{\text{dim}(G)}{2}
		\log\left(\frac{N_FT(R)}{2\pi}\right)
		+\frac{\text{dim}(G)}{2(n-1)}
		\log\det\mathcal K
		+\cdots,
	\end{equation}
	where $S_{A,R}^{(n)}$ is the entropy of a single fermion multiplet.

	To resolve $\mathcal{K}$ further and find its determinant, it is useful to define the correlator of charges between different replicas. The charges are constructed from the chiral currents
	\begin{equation}\label{chiral-currents}
		Q_1^a=\int_{A_1}dx\,J^a(x),
		\qquad
		J_a(x)=:\psi_\alpha^\dagger(x)
		(T_a)_{\alpha\beta}\psi_\beta(x):,
	\end{equation}
	and their replica correlators are correspondingly given by,
	\begin{equation}\label{charge-replica-correlator}
		\mathcal Q_{ij}^{ab}
		=
		\mathcal Q_r^{ab}
		=
		\frac{
			\Tr\!\left(\rho_A^{n-r}Q_1^a
			\rho_A^rQ_1^b\right)}
		{\Tr\!\left(\rho_A^n\right)},
		\qquad r=|i-j|.
	\end{equation}
	Here and below the traces appearing in the definition of $\mathcal Q_r$ are taken in the theory of a single fermion multiplet. The dependence on $r$ follows from the replica-shift symmetry. Let us now relate these correlators to $\mathcal K_{ij}$. The symmetry transformations appearing in the charged moments are
	\begin{equation}
		U_{A_1}(h_i)=e^{iX_i^aQ_1^a},
		\qquad
		U_{A_1}(h_n)
		=
		U_{A_1}(h_{n-1})^{-1}\cdots U_{A_1}(h_1)^{-1}.
	\end{equation}
	The important difference between the last insertion and the others is that $U_{A_1}(h_n)$ depends on all the independent coordinates $X_i^a$. In particular,
	\begin{equation}
		\left.
		\frac{\partial U_{A_1}(h_k)}{\partial X_i^a}
		\right|_{X=0}
		=
		iQ_1^a(\delta_{ik}-\delta_{kn}),
		\qquad i=1,\ldots,n-1.
	\end{equation}
	Taking two derivatives of the charged moments and using the vanishing one-point functions of the charges, we find
	\begin{equation}
		\mathcal K_{ij}^{ab}
		=
		\mathcal Q_{i-j}^{ab}
		-\mathcal Q_i^{ab}
		-\mathcal Q_j^{ab}
		+\mathcal Q_0^{ab}.
	\end{equation}
	The required matrix elements are given by
	\begin{equation}\label{Qij}
		\mathcal Q_{ij}^{ab}
		=
		\int_{A_1}dx\int_{A_1}dy\,
		G_{ij}^{ab}(x,y),
		\qquad
		G_{ij}^{ab}(x,y;A)
		=
		\frac{
			\left\langle\nu J_i^a(x)J_j^b(y)\right\rangle
		}{\langle\nu\rangle}.
	\end{equation}
	For brevity, we defined
	\begin{equation}
		\nu=
		\tau_P(u_1)\tau_{P^{-1}}(v_1)
		\tau_P(u_2)\tau_{P^{-1}}(v_2).
	\end{equation}
	Notice that we are evaluating the correlator at $X_i^a=0$, so the only monodromy involved is the replica permutation \eqref{permutation-matrix}. Since $\nu$ acts trivially on the group indices, from Wick's theorem and definition \eqref{chiral-currents} we can write,
	\begin{equation}
		G_{ij}^{ab}(x,y;A)
		=
		T(R)\delta_{ab}G_{ij}(x,y;A),
	\end{equation}
	where $G_{ij}$ is the corresponding correlator for a single chiral fermion. The remaining calculation is therefore entirely in the replica indices. Writing $\mathcal Q_{ij}^{ab}=T(R)\delta_{ab}\mathcal Q_r$,
	\begin{equation}
		\mathcal K_{ij}
		=
		\mathcal Q_{i-j}
		-\mathcal Q_i
		-\mathcal Q_j
		+\mathcal Q_0,
		\qquad i,j=1,\ldots,n-1.
	\end{equation}
	Arguably the simplest way for understanding the determinant $\det\mathcal K$ is to keep the original $n$ variables $g_i$ of \eqref{charged-moment-NA-n}. Had we done that we would have found a circulant matrix $K_{r}=2\mathcal{Q}_{r}-\mathcal{Q}_{r+1}-\mathcal{Q}_{r-1}$, which is easily diagonalizable by a Fourier sum, but contains a zero mode that corresponds to the fact that in the original $n$ variables the saddle is degenerate. Minding that degeneracy in the saddle expansion one would find,
	\begin{equation}\label{Qm}
		\det\mathcal{K}=\frac{1}{n}\det_{\lambda_m\neq0}K=\frac{1}{n}\prod_{m=1}^{n-1}\left[4\sin^{2}\left(\frac{\pi m}{n}\right)\tilde{\mathcal{Q}}_{m}\right],\qquad\tilde{\mathcal{Q}}_{m}=\sum_{r=1}^{n}\mathcal{Q}_{r}e^{i\frac{2\pi m}{n}r}.
	\end{equation}
	Alternatively, it is also possible to obtain the above through direct diagonalization, however the eigenvectors are far from trivial. Using the identity $\Pi_{m=1}^{n-1}4\sin^{2}\left(\frac{\pi m}{n}\right)=n^{2}$,
	\begin{equation}\label{detQ}
		\det\mathcal K
		=
		n\prod_{m=1}^{n-1}\tilde{\mathcal Q}_m.
	\end{equation}

	We now continue computing $G_{ij}$ and thus $\tilde{\mathcal Q}_m$. Recall the eigenvalues of the permutation matrix $P$ given in \eqref{permutation-matrix} are $e^{2\pi i\theta_m}$, with
	\begin{equation}
		\theta_m=\frac{2m-n-1}{2n},
		\qquad
		S_i^m=\frac{1}{\sqrt n}
		e^{2\pi i\theta_m i},
		\qquad m=1,\ldots,n,
	\end{equation}
	Where $S$ is the basis-transition matrix between the eigenbasis of $P$ and the original replica basis,
	\begin{equation}
		\psi_{i,\alpha}
		=
		\sum_{m=1}^{n}S_i^m\psi_{m,\alpha},
	\end{equation}
	In this basis the twist operators $\nu$ become a product of decoupled vertex operators \eqref{twist-product}, and the currents are now expressed as
	\begin{equation}\label{current-bosonization}
		J_i^a(z)
		=
		\sum_{m,k=1}^{n}
		S_i^{m*}S_i^k
		:\psi_{m,\alpha}^\dagger(z)
		(T_a)_{\alpha\beta}
		\psi_{k,\beta}(z):.
	\end{equation}
	The fields $\psi_{m,\alpha}$ can now be bosonized independently \eqref{bosonoization}.\footnote{We suppress the Klein factors responsible for setting the correct anticommutation relations \cite{vonDelft:1998pk}.} For the fermion bilinears,
	\begin{equation}
		:\psi_m^\dagger(z)\psi_k(z):
		=
		\begin{cases}
			J_m(z), & m=k,\\[2mm]
			\displaystyle
			\frac{1}{2\pi}
			:e^{i\sqrt{4\pi}(\varphi_m(z)-\varphi_k(z))}:,
			& m\neq k.
		\end{cases}
	\end{equation}
	It will be convenient to define the correlators
	\begin{equation}
		G_{mk}^{(1)}
		=
		\frac{\langle\nu J_m(x)J_k(y)\rangle}
		{\langle\nu\rangle},
		\qquad
		G_{m,kl}^{(2)}
		=
		\frac{
			\left\langle
			\nu J_m(x)
			:e^{i\sqrt{4\pi}(\varphi_k(y)-\varphi_l(y))}:
			\right\rangle}
		{\langle\nu\rangle},
	\end{equation}
	\begin{equation}
		G_{mk,lp}^{(3)}
		=
		\frac{
			\left\langle
			\nu
			:e^{i\sqrt{4\pi}(\varphi_m(x)-\varphi_k(x))}:
			:e^{i\sqrt{4\pi}(\varphi_l(y)-\varphi_p(y))}:
			\right\rangle}
		{\langle\nu\rangle}.
	\end{equation}
	Through a standard computation in the bosonic theory we find,
	\begin{equation}\label{helpful-correlators}
		G_{mk}^{(1)}=\frac{1}{(2\pi)^{2}}\left(\frac{\theta_{m}\theta_{k}P'(x)P'(y)}{P(x)P(y)}+\frac{\delta_{mk}}{|x-y|^{2}}\right),\qquad G_{m,kl}^{(2)}=\delta_{kl}\frac{\langle\nu J_{m}\rangle}{\langle\nu\rangle},
	\end{equation}
	\begin{equation}\label{helpful-correlators2}
		G_{mk,lp}^{(3)}=\left(\frac{P(x)}{P(y)}\right)^{\theta_{m}-\theta_{k}}|x-y|^{2\delta_{mk}-2}\delta_{mp}\delta_{kl}.
	\end{equation}
	Where we defined
	\begin{equation}
		P(x)
		=
		\frac{|x-u_1||x-u_2|}
		{|x-v_1||x-v_2|}.
	\end{equation}

	Using \eqref{current-bosonization}, \eqref{helpful-correlators}, \eqref{helpful-correlators2}, and the explicit form of $S_i^m$, we find the correlator in \eqref{Qij},
	\begin{equation}
		G_{ij}(x,y;A)=\frac{1}{(2\pi)^{2}}\frac{1}{|x-y|^{2}}\frac{1}{n^{2}}\frac{\left(P\!\left(x\right)-P\!\left(y\right)\right)^{2}}{4P\!\left(x\right)P\!\left(y\right)}\frac{1}{\sinh^{2}\frac{1}{2n}\left(\log\frac{P(x)}{P(y)}+i2\pi(i-j)\right)}.
	\end{equation}
	The replica-shift symmetry $G_{ij}=G_{|i-j|}$ is clearly seen. Given that the determinant depends on the Fourier modes \eqref{Qm}, it is convenient to define
	\begin{equation} \label{G_m}
		\tilde{G}_{m}(x,y)=\sum_{r=1}^{n}G_{r}(x,y)e^{i\frac{2\pi m}{n}r}=\frac{1}{(2\pi)^{2}}\frac{1}{|x-y|^{2}}\left[\frac{n-m}{n}\left(\frac{P(x)}{P(y)}\right)^{-\frac{m}{n}}+\frac{m}{n}\left(\frac{P(x)}{P(y)}\right)^{\frac{n-m}{n}}\right].
	\end{equation}
	Performing the double integral over $G_m$ to get the charge correlator yields,\footnote{
	Using $[Q_A,\rho_A]=0$ and $Q_A=Q_1+Q_2$, the nonzero replica Fourier modes satisfy ${\tilde{\mathcal{Q}}_{m}^{(A_{1},A_{1})}=-\tilde{\mathcal{Q}}_{m}^{(A_{1},A_{2})}}$. We can therefore perform the integral over $A_1\times A_2$. Mapping the endpoints to $(0,\eta,1,\infty)$ and introducing $x=\eta t$, $y=1/u$, $s=m/n$, and $f(t)=t(1-\eta t)/(1-t)$, an integration by parts gives \[\tilde{\mathcal Q}_m=\frac{\eta(1-s)}{4\pi^2}\left(\int_0^1dt\,\frac{f(t)^{-s}}{1-t}\right)\left(\int_0^1du\,f(u)^{-s}\right).\] The two integrals are Euler representations of ${}_2F_1(s,1-s;1;\eta)$ and ${}_2F_1(s,1-s;2;\eta)$, respectively.}
	
	\begin{equation}
		\tilde{\mathcal Q}_m
		=
		\frac{\eta}{4}
		\frac{
			\frac mn\left(\frac{n-m}{n}\right)
		}{
			\sin^2\left(\frac{\pi m}{n}\right)
		}
		{}_2F_1\left(
		\frac mn,\frac{n-m}{n};1;\eta
		\right)
		{}_2F_1\left(
		\frac mn,\frac{n-m}{n};2;\eta
		\right).
	\end{equation}
	From \eqref{detQ}, we therefore find
	\begin{equation}
		\det\mathcal K
		=
		\eta^{n-1}
		\frac{\Gamma(n)^2}{n^{2n-1}}
		\prod_{m=1}^{n-1}
		{}_2F_1\left(
		\frac mn,\frac{n-m}{n};1;\eta
		\right)
		{}_2F_1\left(
		\frac mn,\frac{n-m}{n};2;\eta
		\right).
	\end{equation}
	In conclusion, we obtain the charged moments at leading order in large $N_F$,
	\begin{align} \label{charged-moment-largeN}
		\frac{\Tr(\tilde\rho_D^n)}{\Tr(\rho_A^n)}
		\approx{}&
		\frac{1}{|G|^{n-1}}
		\left(\frac{2\pi}{N_FT(R)}\right)^{
			\frac{(n-1)\text{dim}(G)}{2}}
		\nonumber\\
		&\times
		\left[
		\frac{n^{2n-1}}
		{\eta^{n-1}\Gamma(n)^2}
		\prod_{m=1}^{n-1}
		\left(
		{}_2F_1\left(
		\frac mn,\frac{n-m}{n};1;\eta
		\right)
		{}_2F_1\left(
		\frac mn,\frac{n-m}{n};2;\eta
		\right)
		\right)^{-1}
		\right]^{\frac{\text{dim}(G)}{2}}.
	\end{align}
	The analytic continuation to $n\rightarrow1$ can be performed using the Sommerfeld--Watson method, following the calculation in \eqref{fermion_SW_EOP},
	
	\begin{equation}
		\begin{aligned}
			\widetilde S_A-&N_F S_{A,R}
			=\log|G|
			+\frac{\dim(G)}{2}\Bigg(\log\frac{N_{F}T(R)}{2\pi}-2\gamma_{E}-1\Bigg)\\
			&\quad+\frac{\dim(G)}{2}\int_{0}^{\infty}\frac{\pi\,ds}{\cosh^{2}(\pi s)}\log\left[\eta\,{}_{2}F_{1}\left(\frac{1}{2}+is,\frac{1}{2}-is;1;\eta\right){}_{2}F_{1}\left(\frac{1}{2}+is,\frac{1}{2}-is;2;\eta\right)\right].
		\end{aligned}
	\end{equation}

	Let us analyze the result by considering the nearly touching intervals limit $\eta\rightarrow1$,
	\begin{equation}
		\tilde{\mathcal Q}_m
		=
		-\frac{1}{(2\pi)^2}\log(1-\eta)
		+\mathcal O(1).
	\end{equation}
	The familiar universal behavior follows,
	\begin{equation}
		\tilde I_A^{(n)}-N_FI_{A,R}^{(n)}
		=
		-
		\frac{\text{dim}(G)}{2}
		\log\left[
		-N_FT(R)\log(1-\eta)
		\right]
		+\mathcal O(1).
	\end{equation}
	
	To conclude, we notice that the saddle expansion is unreliable in the opposite limit of large interval separation $\eta\ll1$. This is because the holonomy dependence of $\log Z_n(h_i,0)$ decays in this limit, and consequently $\mathcal K$ approaches zero. The physical limit relevant for large separation is therefore the opposite one: fixed $N_F$ (possibly of order one), followed by $\eta\rightarrow0$.	Even so, we can retrieve the  $\eta\ll1$ asymptotics by simply using the complementarity relation \eqref{complementary-duality}, the result \eqref{NA-b-cycle-average}, and the factorization \eqref{saddle},
	\begin{equation}
		\frac{\Tr(\tilde\rho_D^n)}{\Tr(\rho_A^n)}
		=
		\int\frac{d^{n-1}h}{|G|^{n-1}}
		\left[
		\frac{Z_n(h_i,0)}{Z_n(0,0)}
		\right]^{N_F}
		=
		1-\frac{N_Fd_R(n^2-1)}{12n}\eta
		+\mathcal O(\eta^2).
	\end{equation}\\

	\section{Discussion}
	We studied the entropic order parameter using the replica trick, and found exact solutions through averages over twisted partition functions. Moreover, we showed that the complete averaging over all non-contractible cycles gives access to additional  data, which is the decrease in Holevo information of complementary-charge measurement that occurs once removing the intertwiners. Curiously, we found that for the Dirac fermion it is equal to the difference between the mutual information of the $\mathbb{Z}^F_2$-gauged and ungauged theories, independently of which even $\mathbb Z_N\!\subset\!U(1)$ subgroup is considered. Additionally, it vanishes for the compact boson. Largely, it appears that the Holevo decrease is more robust than the EOP, and it will be important to check if this is unique to free theories or holds more generally. A larger question is whether the relation with gauging is more general. The fact that the double average involves a summation over all twist sectors suggests a possible connection. To investigate this further, it would be interesting to consider other theories, such as the Schwinger model.

	Another promising direction that we plan to pursue is to study anomalous symmetries. The extension of the formalism is potentially connected to the entropic certainty relation introduced in \cite{Casini2020}, which we would like to understand in terms of the replica-trick. Finally, it would be interesting to study higher dimensions. The recent work of \cite{Abate:2026yzt} may be a promising direction to that end.

 	\section*{Acknowledgments}
	I would like to thank Horacio Casini, Valentín Benedetti, and Javier Magan for helpful discussions and feedback. The research of T.S. is supported by the Simons-Balseiro fellowship grant. The author acknowledges the use of ChatGPT (OpenAI) for checking intermediate calculations, editing suggestions, and figure generation. All results and conclusions were independently verified by the author.
	
	\appendix
	\section{Theta Function Averages}
	\label{APP-theta-identities}
	
	This appendix details the group averages of the Riemann theta functions \eqref{riemann_theta_characteristic} and its modulus square required for the Dirac fermion, as well the positivity claimed for \eqref{fermion_holevo_theta}.
	
	\begin{itemize}
		\item The $U\!\left(1\right)$ averages for $\Theta,|\Theta|^2$ are respectively given in subsections \ref{app-U1Theta}, \ref{app-U1Thetasq}.
		
		\item The $\mathbb{Z}_N$ averages for $\Theta,|\Theta|^2$ are respectively given in subsections \ref{app-ZNTheta}, \ref{app-ZNThetasq}.
		
		\item Positivity of $\Theta(0|2\Gamma_{n})\Theta(0|\frac{\Gamma_{n}}{2})/\Theta(0|\Gamma_{n})^{2}$. is proved in subsection \ref{app-thetapos}.
	\end{itemize}
	For convenience, we recall that after setting $\tilde{\boldsymbol\alpha}=\boldsymbol\alpha/(2\pi)$ and $\tilde{\boldsymbol\beta}=\boldsymbol\beta/(2\pi)$, the holonomy dependence of the twisted partition functions is encoded in the Riemann theta function with characteristics,
	\begin{equation}\label{theta_holonomy_expansion}
		\Theta\!\begin{bmatrix}
			\tilde{\boldsymbol\beta}\\ \tilde{\boldsymbol\alpha}
		\end{bmatrix}(0|\Gamma_n)
		=\sum_{\boldsymbol m\in\mathbb Z^{n-1}}
		e^{i\pi(\boldsymbol m+\tilde{\boldsymbol\beta})^{\rm T}
			\Gamma_n(\boldsymbol m+\tilde{\boldsymbol\beta})
			+2\pi i(\boldsymbol m+\tilde{\boldsymbol\beta})^{\rm T}
			\tilde{\boldsymbol\alpha}}.
	\end{equation}
	
	\hiddensubsection{$U\!\left(1\right)$ Average: $\Theta$} \label{app-U1Theta}
	\paragraph*{$a$-cycles.} First, setting $\boldsymbol\beta=0$ and integrating over the $a$-cycle holonomies gives
	\begin{equation}\label{theta_alpha_average}
		\int\frac{d^{n-1}\alpha}{(2\pi)^{n-1}}
		\Theta\!\begin{bmatrix}
			\boldsymbol0\\ \tilde{\boldsymbol\alpha}
		\end{bmatrix}(0|\Gamma_n)
		=\sum_{\boldsymbol m\in\mathbb Z^{n-1}}
		e^{i\pi\boldsymbol m^{\rm T}\Gamma_n\boldsymbol m}
		\prod_{j=1}^{n-1}\delta_{m_j,0}
		=1.
	\end{equation}
	\paragraph*{$b$-cycles.} Setting $\boldsymbol\alpha=0$ and integrating over the $b$-cycle holonomies unfolds the sum into a Gaussian integral over $\mathbb{R}^{n-1}$,
	\begin{equation}\label{theta_gaussian_unfolding}
		\begin{aligned}
			&\int\frac{d^{n-1}\beta}{(2\pi)^{n-1}}
			\Theta\!\begin{bmatrix}
				\tilde{\boldsymbol\beta}\\ \boldsymbol0
			\end{bmatrix}(0|\Gamma_n)
			=\sum_{\boldsymbol m\in\mathbb Z^{n-1}}
			\int_{[-\frac{1}{2},\frac{1}{2}]^{n-1}}d^{n-1}\tilde\beta\,
			e^{i\pi(\boldsymbol m+\tilde{\boldsymbol\beta})^{\rm T}
				\Gamma_n(\boldsymbol m+\tilde{\boldsymbol\beta})}\\
			&\quad=
			\sum_{\boldsymbol m\in\mathbb Z^{n-1}}
			\int_{\prod_{j=1}^{n-1}[m_j-\frac{1}{2},m_j+\frac{1}{2}]}d^{n-1}x\,
			e^{i\pi\boldsymbol x^{\rm T}\Gamma_n\boldsymbol x}
			=\int_{\mathbb R^{n-1}}d^{n-1}x\,
			e^{i\pi\boldsymbol x^{\rm T}\Gamma_n\boldsymbol x}
			=\frac{1}{\sqrt{\det(-i\Gamma_n)}}.
		\end{aligned}
	\end{equation}
	Here we used that $\Gamma_n$ is purely imaginary, with positive-definite imaginary part. Consequently,
	\begin{equation}\label{theta_b_gaussian}
		\int\frac{d^{n-1}\beta}{(2\pi)^{n-1}}\frac{Z_{n}(0,\beta_{i})}{Z_{n}(0,0)}=\frac{1}{\sqrt{\det(-i\Gamma_{n})}\,\Theta(0|\Gamma_{n})}=\frac{1}{\Theta(0|-\Gamma_{n}^{-1})},
	\end{equation}
	where the last equality follows from the Poisson-resummation identity,
	\begin{equation}\label{theta_modular_boson}
		\Theta(0|\Gamma_{n})=\frac{\Theta(0|-\Gamma_{n}^{-1})}{\sqrt{\det(-i\Gamma_{n})}}.
	\end{equation}
	\hiddensubsection{$U\!\left(1\right)$ Average: $|\Theta|^2$} \label{app-U1Thetasq}
	Since $\Gamma_n$ is purely imaginary, there is a little simplification,
	\begin{equation}\label{dirac_theta_double_sum}
			\left|\Theta\!\begin{bmatrix}
				\tilde{\boldsymbol\beta}\\ \tilde{\boldsymbol\alpha}
			\end{bmatrix}(0|\Gamma_n)\right|^2
			=			
			\sum_{\boldsymbol m,\boldsymbol m'\in\mathbb Z^{n-1}}
			e^{i\pi(\boldsymbol m+\tilde{\boldsymbol\beta})^{\rm T}
				\Gamma_n(\boldsymbol m+\tilde{\boldsymbol\beta})
				+i\pi(\boldsymbol m'+\tilde{\boldsymbol\beta})^{\rm T}
				\Gamma_n(\boldsymbol m'+\tilde{\boldsymbol\beta})
				+2\pi i(\boldsymbol m-\boldsymbol m')^{\rm T}
				\tilde{\boldsymbol\alpha}}.
	\end{equation}
	\paragraph*{$a$-cycles.} Setting $\boldsymbol\beta=0$ and averaging over $\boldsymbol\alpha$ imposes $\boldsymbol m=\boldsymbol m'$, giving
	\begin{equation} \label{thetasqr_alpha_average}
		\int\frac{d^{n-1}\alpha}{(2\pi)^{n-1}}
		\left|\Theta\!\begin{bmatrix}
			\boldsymbol0\\ \tilde{\boldsymbol\alpha}
		\end{bmatrix}(0|\Gamma_n)\right|^2
		=\sum_{\boldsymbol m\in\mathbb Z^{n-1}}
		e^{2\pi i\boldsymbol m^{\rm T}\Gamma_n\boldsymbol m}
		=\Theta(0|2\Gamma_n).
	\end{equation}
	\paragraph*{$b$-cycles.} 
	Set $\boldsymbol\alpha=0$ and introduce $\boldsymbol d=\boldsymbol m-\boldsymbol m'$. For each fixed $\boldsymbol d$, unfolding the sum over $\boldsymbol m'$ into the integral over $\tilde{\boldsymbol\beta}$ gives
	\begin{equation}\label{thetasqr_b_gaussian}
		\begin{aligned}
			&\int\frac{d^{n-1}\beta}{(2\pi)^{n-1}}
			\left|\Theta\!\begin{bmatrix}
				\tilde{\boldsymbol\beta}\\ \boldsymbol0
			\end{bmatrix}(0|\Gamma_n)\right|^2
			=
			\sum_{\boldsymbol d\in\mathbb Z^{n-1}}
			e^{\frac{i\pi}{2}\boldsymbol d^{\rm T}
				\Gamma_n\boldsymbol d}
			\int_{\mathbb R^{n-1}}d^{n-1}\tilde\beta\,
			e^{2\pi i(\tilde{\boldsymbol\beta}+\boldsymbol d/2)^{\rm T}
				\Gamma_n(\tilde{\boldsymbol\beta}+\boldsymbol d/2)}
			\\
			&\quad=
			\frac{\Theta(0|\Gamma_n/2)}
			{\sqrt{\det(-2i\Gamma_n)}}.
		\end{aligned}
	\end{equation}
	
	\paragraph*{$a,b$-cycles.} Averaging over both cycle families imposes $\boldsymbol m=\boldsymbol m'$ and unfolds the remaining sum,
	\begin{equation} \label{thetasqr_ab_gaussian}
		\int\frac{d^{n-1}\alpha\,d^{n-1}\beta}{(2\pi)^{2(n-1)}}
		\left|\Theta\!\begin{bmatrix}
			\tilde{\boldsymbol\beta}\\ \tilde{\boldsymbol\alpha}
		\end{bmatrix}(0|\Gamma_n)\right|^2
		=\int_{\mathbb R^{n-1}}d^{n-1}x\,
		e^{2\pi i\boldsymbol x^{\rm T}\Gamma_n\boldsymbol x}
		=\frac{1}{\sqrt{\det(-2i\Gamma_n)}}.
	\end{equation}
	Dividing by the untwisted partition function $Z_n^{(D)}(0,0)\propto\Theta(0|\Gamma_n)^2$, we recover \eqref{fermion_dirac_integrated}.

	\hiddensubsection{$\mathbb Z_N$ Average: $\Theta$} \label{app-ZNTheta}
	\paragraph*{$a$-cycles.} The main difference is that now equalities between indices are modulo $N$, the $a$-cycle average gives
	\begin{equation}\label{theta_ZN_a_average}
		\begin{aligned}
			&\frac{1}{N^{n-1}}\sum_{\boldsymbol s\in\mathbb Z_N^{n-1}}
			\Theta\!\begin{bmatrix}\boldsymbol0\\ \boldsymbol s/N\end{bmatrix}(0|\Gamma_n)
			=\sum_{\boldsymbol m\in\mathbb Z^{n-1}}
			e^{i\pi\boldsymbol m^{\rm T}\Gamma_n\boldsymbol m}
			\prod_{j=1}^{n-1}\delta_{m_j,0}^{[N]}\\
			&\qquad=\sum_{\boldsymbol k\in\mathbb Z^{n-1}}
			e^{i\pi N^2\boldsymbol k^{\rm T}\Gamma_n\boldsymbol k}
			=\Theta(0|N^2\Gamma_n).
		\end{aligned}
	\end{equation}
	Here $\delta_{m,0}^{[N]}$ denotes the Kronecker delta modulo $N$.\footnote{Notice the normalization $\frac{1}{N^{n-1}}\sum_{\boldsymbol s\in\mathbb Z_N^{n-1}}e^{\frac{2\pi i}{N}\boldsymbol s^{\rm T}\boldsymbol d}=\delta_{\boldsymbol d,\boldsymbol 0}^{[N]}$.}
	\paragraph*{$b$-cycles.} The $b$-cycle average becomes a sum over a finer lattice with spacing $N^{-1}$: the pairs $(\boldsymbol m,\boldsymbol s)\in\mathbb Z^{n-1}\times\mathbb Z_N^{n-1}$ uniquely define points of the finer lattice $N^{-1}\mathbb Z^{n-1}$ through $\boldsymbol m+\boldsymbol s/N$. Hence,
	\begin{equation} \label{theta_ZN_b_average}
		\begin{aligned}
			&\frac{1}{N^{n-1}}\sum_{\boldsymbol s\in\mathbb Z_N^{n-1}}
			\Theta\!\begin{bmatrix}\boldsymbol s/N\\ \boldsymbol0\end{bmatrix}(0|\Gamma_n)
			=\frac{1}{N^{n-1}}\sum_{\boldsymbol m\in\mathbb Z^{n-1}}
			\sum_{\boldsymbol s\in\mathbb Z_N^{n-1}}
			e^{i\pi(\boldsymbol m+\frac{\boldsymbol{s}}{N})^{\rm T}
				\Gamma_n(\boldsymbol m+\frac{\boldsymbol{s}}{N})}\ \\
			&\qquad=\frac{1}{N^{n-1}}\sum_{\boldsymbol k\in\mathbb Z^{n-1}}
			e^{\frac{i\pi}{N^2}\boldsymbol k^{\rm T}\Gamma_n\boldsymbol k}
			=\frac{\Theta(0|\Gamma_n/N^2)}{N^{n-1}}.
		\end{aligned}
	\end{equation}
	\hiddensubsection{$\mathbb Z_N$ Average: $|\Theta|^2$} \label{app-ZNThetasq}
	\paragraph*{$a$-cycles.} The average over \eqref{dirac_theta_double_sum} now imposes $\boldsymbol m-\boldsymbol m'=N\boldsymbol k$, with $\boldsymbol k\in\mathbb Z^{n-1}$. Introducing $\boldsymbol u=(\boldsymbol m+\boldsymbol m')/2$ and changing summation variables gives
	\begin{equation}\label{dirac_ZN_a_lattice}
		\frac{1}{N^{n-1}}\sum_{\boldsymbol s\in\mathbb Z_N^{n-1}}
		\left|\Theta\!\begin{bmatrix}
			\boldsymbol0\\ \boldsymbol s/N
		\end{bmatrix}(0|\Gamma_n)\right|^2
		=
		\sum_{\boldsymbol k\in\mathbb Z^{n-1}}
		e^{\frac{i\pi N^2}{2}\boldsymbol k^{\rm T}\Gamma_n\boldsymbol k}
		\sum_{\boldsymbol u\in\tilde{\mathbb Z}_a^{n-1}(\boldsymbol k)}
		e^{2\pi i\boldsymbol u^{\rm T}\Gamma_n\boldsymbol u}.
	\end{equation}
	The shifted lattice $\tilde{\mathbb{Z}}^{n-1}(\boldsymbol{k})=\mathbb{Z}^{n-1}+\frac{N}{2}\boldsymbol{k}$ is simply equal to $\mathbb{Z}^{n-1}$ for $N$ even. In that case, the two sums factorize,
	\begin{equation}\label{dirac_ZN_a_even}
		\frac{1}{N^{n-1}}\sum_{\boldsymbol s}
		\left|\Theta\!\begin{bmatrix}
			\boldsymbol0\\ \boldsymbol s/N
		\end{bmatrix}(0|\Gamma_n)\right|^2
		=\Theta(0|N^2\Gamma_n/2)\Theta(0|2\Gamma_n),
		\qquad N\ \mathrm{even}.
	\end{equation}
	For $N$ odd, $\tilde{\mathbb{Z}}^{n-1}(\boldsymbol{k})$ depends on the parity class of $\boldsymbol k$. Namely, decomposing $\boldsymbol k=2\boldsymbol\ell+\boldsymbol\epsilon$, with $\boldsymbol\epsilon\in\{0,1\}^{n-1}$, gives
	\begin{equation}\label{dirac_ZN_a_odd}
		\frac{1}{N^{n-1}}\sum_{\boldsymbol s}
		\left|\Theta\!\begin{bmatrix}
			\boldsymbol0\\ \boldsymbol s/N
		\end{bmatrix}(0|\Gamma_n)\right|^2
		=
		\sum_{\boldsymbol\epsilon\in\{0,1\}^{n-1}}
		\Theta\!\begin{bmatrix}\boldsymbol\epsilon/2\\ \boldsymbol0\end{bmatrix}(0|2N^2\Gamma_n)
		\Theta\!\begin{bmatrix}\boldsymbol\epsilon/2\\ \boldsymbol0\end{bmatrix}(0|2\Gamma_n),
		\qquad N\ \mathrm{odd}.
	\end{equation}
	\paragraph*{$b$-cycles.}
	For the $b$-cycle average, we introduce $\boldsymbol d=\boldsymbol m-\boldsymbol m'$ and the holonomy-shifted midpoint $\boldsymbol u=(\boldsymbol m+\boldsymbol m')/2+\boldsymbol s/N$. As in the single-theta average, the definition $\boldsymbol k=N\boldsymbol m'+\boldsymbol s$ absorbs $\boldsymbol m'$ and the discrete holonomy into a single integer vector, so that $\boldsymbol u=\boldsymbol k/N+\boldsymbol d/2$. Averaging \eqref{dirac_theta_double_sum},
	\begin{equation}\label{dirac_ZN_b_lattice}
			\frac{1}{N^{n-1}}\sum_{\boldsymbol s\in\mathbb Z_N^{n-1}}
			\left|\Theta\!\begin{bmatrix}
				\boldsymbol s/N\\ \boldsymbol0
			\end{bmatrix}(0|\Gamma_n)\right|^2
			=
			\frac{1}{N^{n-1}}
			\sum_{\boldsymbol d\in\mathbb Z^{n-1}}
			e^{\frac{i\pi}{2}\boldsymbol d^{\rm T}\Gamma_n\boldsymbol d}
			\sum_{\boldsymbol u\in\tilde{\mathbb{Z}}_{1/N}^{n-1}(\boldsymbol d)}
			e^{2\pi i\boldsymbol u^{\rm T}\Gamma_n\boldsymbol u}.
	\end{equation}
	The shifted lattice $\tilde{\mathbb{Z}}_{1/N}^{n-1}(\boldsymbol d)=N^{-1}\mathbb Z^{n-1}+\boldsymbol d/2$ is simply equal to $N^{-1}\mathbb Z^{n-1}$ for $N$ even. In that case, the two sums factorize as in the other average,
	\begin{equation}\label{dirac_ZN_b_even}
		\frac{1}{N^{n-1}}\sum_{\boldsymbol s}
		\left|\Theta\!\begin{bmatrix}
			\boldsymbol s/N\\ \boldsymbol0
		\end{bmatrix}(0|\Gamma_n)\right|^2
		=\frac{\Theta(0|\Gamma_n/2)\Theta(0|2\Gamma_n/N^2)}{N^{n-1}},
		\qquad N\ \mathrm{even}.
	\end{equation}
	For $N$ odd, $\tilde{\mathbb{Z}}_{1/N}^{n-1}(\boldsymbol d)$ depends on the parity class of $\boldsymbol d$. Namely, decomposing $\boldsymbol d=2\boldsymbol\ell+\boldsymbol\epsilon$, with $\boldsymbol\epsilon\in\{0,1\}^{n-1}$, gives
	\begin{equation}\label{dirac_ZN_b_odd}
		\frac{1}{N^{n-1}}\sum_{\boldsymbol s}
		\left|\Theta\!\begin{bmatrix}
			\boldsymbol s/N\\ \boldsymbol0
		\end{bmatrix}(0|\Gamma_n)\right|^2
		=
		\frac{1}{N^{n-1}}
		\sum_{\boldsymbol\epsilon\in\{0,1\}^{n-1}}
		\Theta\!\begin{bmatrix}\boldsymbol\epsilon/2\\ \boldsymbol0\end{bmatrix}(0|2\Gamma_n)
		\Theta\!\begin{bmatrix}\boldsymbol\epsilon/2\\ \boldsymbol0\end{bmatrix}(0|2\Gamma_n/N^2),
		\qquad N\ \mathrm{odd}.
	\end{equation}
	\paragraph*{$a,b$-cycles.}
	Finally, averaging over both cycle families combines the two restrictions. The discrete $a$-cycle average imposes $\boldsymbol m-\boldsymbol m'=N\boldsymbol\ell$, with $\boldsymbol\ell\in\mathbb Z^{n-1}$, while the $b$-cycle average refines the lattice of the holonomy-shifted midpoint $\boldsymbol u=(\boldsymbol m+\boldsymbol m')/2+\boldsymbol t/N$. Thus,
	\begin{equation}\label{dirac_ZN_double_lattice}
			\frac{1}{N^{2(n-1)}}\sum_{\boldsymbol s,\boldsymbol t\in\mathbb Z_N^{n-1}}
			\left|\Theta\!\begin{bmatrix}
				\boldsymbol t/N\\ \boldsymbol s/N
			\end{bmatrix}(0|\Gamma_n)\right|^2
			=
			\frac{1}{N^{n-1}}
			\sum_{\boldsymbol\ell\in\mathbb Z^{n-1}}
			e^{\frac{i\pi N^2}{2}\boldsymbol\ell^{\rm T}\Gamma_n\boldsymbol\ell}
			\sum_{\boldsymbol u\in\tilde{\mathbb{Z}}_{1/N}^{n-1}(\boldsymbol\ell)}
			e^{2\pi i\boldsymbol u^{\rm T}\Gamma_n\boldsymbol u}.
	\end{equation}
	The shifted lattice $\tilde{\mathbb{Z}}_{1/N}^{n-1}(\boldsymbol\ell)=N^{-1}\mathbb Z^{n-1}+\frac{N}{2}\boldsymbol\ell$ is simply equal to $N^{-1}\mathbb Z^{n-1}$ for $N$ even. In that case, the two sums factorize,
	\begin{equation}\label{dirac_ZN_double_even}
		\frac{1}{N^{2(n-1)}}\sum_{\boldsymbol s,\boldsymbol t}
		\left|\Theta\!\begin{bmatrix}
			\boldsymbol t/N\\ \boldsymbol s/N
		\end{bmatrix}(0|\Gamma_n)\right|^2
		=\frac{\Theta(0|N^2\Gamma_n/2)\Theta(0|2\Gamma_n/N^2)}{N^{n-1}},
		\qquad N\ \mathrm{even}.
	\end{equation}
	For $N$ odd, $\tilde{\mathbb{Z}}_{1/N}^{n-1}(\boldsymbol\ell)$ depends on the parity class of $\boldsymbol\ell$. Namely, decomposing $\boldsymbol\ell=2\boldsymbol p+\boldsymbol\epsilon$, with $\boldsymbol\epsilon\in\{0,1\}^{n-1}$, gives
	\begin{equation}\label{dirac_ZN_double_odd}
		\begin{aligned}
			&\frac{1}{N^{2(n-1)}}\sum_{\boldsymbol s,\boldsymbol t}
			\left|\Theta\!\begin{bmatrix}
				\boldsymbol t/N\\ \boldsymbol s/N
			\end{bmatrix}(0|\Gamma_n)\right|^2\\
			&\qquad=
			\frac{1}{N^{n-1}}
			\sum_{\boldsymbol\epsilon\in\{0,1\}^{n-1}}
			\Theta\!\begin{bmatrix}\boldsymbol\epsilon/2\\ \boldsymbol0\end{bmatrix}(0|2N^2\Gamma_n)
			\Theta\!\begin{bmatrix}\boldsymbol\epsilon/2\\ \boldsymbol0\end{bmatrix}(0|2\Gamma_n/N^2),
			\qquad N\ \mathrm{odd}.
		\end{aligned}
	\end{equation}
	The normalized partition-function averages follow by dividing these expressions by $\Theta(0|\Gamma_n)^2$. In particular, for even $N$, all $N$-dependent factors cancel in the replica combination \eqref{Holevo-diff-replica} entering the Holevo-information difference \eqref{ZN-Holevo-even}.
	
	\hiddensubsection{Positivity of the Theta-Function Ratio} \label{app-thetapos}
	\label{APP-theta-positivity}
	We show that $\Theta(0|2\Gamma_n)\Theta(0|\Gamma_n/2)\geq\Theta(0|\Gamma_n)^2$ for the purely imaginary period matrix $\Gamma_n=iY_n$. Write $\boldsymbol{m}=\boldsymbol{2k}+\boldsymbol{\epsilon}$, where $\boldsymbol\epsilon\in\{0,1\}^{n-1}$. Each point on the lattice $\mathbb{Z}^{n-1}$ is uniquely determined by $\boldsymbol{k},\boldsymbol{\epsilon}$. Define
	\begin{equation}
		T_{\boldsymbol\epsilon}
		=
		\sum_{\boldsymbol k\in\mathbb Z^{n-1}}
		e^{-2\pi(\boldsymbol k+\boldsymbol\epsilon/2)^{\rm T}
			Y_n(\boldsymbol k+\boldsymbol\epsilon/2)}.
	\end{equation}
	Splitting the integer sums into their parity sectors gives
	\begin{equation}
		\Theta(0|\Gamma_n/2)
		=\sum_{\boldsymbol\epsilon}T_{\boldsymbol\epsilon},
		\qquad
		\Theta(0|\Gamma_n)^2
		=\sum_{\boldsymbol\epsilon}T_{\boldsymbol\epsilon}^2,
		\qquad
		\Theta(0|2\Gamma_n)=T_{\boldsymbol0}.
	\end{equation}
	It can be proven by Poisson resummation that $T_{\boldsymbol\epsilon}\leq T_{\boldsymbol0}$, \cite{Regev2017}. Consequently,
	\begin{equation}
		\Theta(0|2\Gamma_n)\Theta(0|\Gamma_n/2)
		-\Theta(0|\Gamma_n)^2
		=
		\sum_{\boldsymbol\epsilon}
		T_{\boldsymbol\epsilon}
		(T_{\boldsymbol0}-T_{\boldsymbol\epsilon})
		\geq0.
	\end{equation}

	\section{Compact Boson: Winding Sectors and Classical Action}
	\label{APP-boson-sectors}
	
	We briefly review how the classical winding sectors of the compact boson are defined on a Riemann surface and compute their classical action. Let $\Sigma=\mathcal R_n$ be a compact Riemann surface of genus $g=n-1$. There are $2g$ independent non-contractible cycles, and we define a canonical homology basis by the set of cycles $\{a_i,b_i\}_{i=1}^{g}$ satisfying $a_i\circ a_j=b_i\circ b_j=0,\; a_i\circ b_j=\delta_{ij}$, where $\circ$ denotes the oriented intersection number. The compact field $\phi\sim\phi+2\pi R$ need not be single-valued around these cycles,
	\begin{equation}\label{boson_periods}
		\oint_{a_i}d\phi_{\rm cl}=2\pi R\left(m_i+\frac{\beta_i}{2\pi}\right),
		\qquad
		\oint_{b_i}d\phi_{\rm cl}=2\pi R\left(n_i+\frac{\alpha_i}{2\pi}\right),
		\qquad m_i,n_i\in\mathbb Z,
	\end{equation}
	where $\alpha_i$ and $\beta_i$ are the holonomies, defined modulo
	$2\pi$. Notice that the insertion of a symmetry defect across $a_i$ imposes a holonomy on the dual cycle $b_i$, and vice-versa.
	
	To construct the classical solutions, recall some elementary facts about one-forms on a Riemann surface. In a local complex coordinate $z=x+iy$, a complex one-form can be decomposed into its $(1,0)$ and $(0,1)$ components,
	\begin{equation}
		\xi=f(z,\bar z)\,dz+g(z,\bar z)\,d\bar z.
	\end{equation}
	A holomorphic one-form is locally of the form $\omega=f(z)\,dz$. Although $dz$ need not be globally defined on a higher-genus surface, the combinations $f(z)\,dz$ patch together to define global holomorphic one-forms. There are precisely $g$ linearly independent ones which we denote by $\omega_i$. Equivalently, the complex harmonic one-forms admit the Hodge decomposition 
	$H^1(\Sigma,\mathbb C)=H^{1,0}(\Sigma)\oplus H^{0,1}(\Sigma)$. Their complex conjugates $\bar\omega_i$ span the antiholomorphic one-forms. In particular, every real harmonic one-form such as $d\phi$ can be written as a linear combination of $\omega_i$ and $\bar\omega_i$ with complex-conjugate coefficients.
	
	We normalize the holomorphic forms by
	\begin{equation}
		\oint_{a_i}\omega_j=\delta_{ij},
		\qquad
		\oint_{b_i}\omega_j=\Gamma_{ij},
	\end{equation}
	The period matrix $\Gamma=X+iY$ is symmetric and $Y=\Im\Gamma$ is positive definite. We also introduce the real harmonic forms $\mathsf a_i,\mathsf b_i$ dual to the cycles,
	\begin{equation}\label{real_harmonic_basis}
		\oint_{a_i}\mathsf a_j=\oint_{b_i}\mathsf b_j=\delta_{ij},
		\qquad
		\oint_{a_i}\mathsf b_j=\oint_{b_i}\mathsf a_j=0.
	\end{equation}
	These real forms can be constructed explicitly from the holomorphic
	basis. Write
	\begin{equation}
		\mathsf a_i=A_{ij}\omega_j+\bar A_{ij}\bar\omega_j,
		\qquad
		\mathsf b_i=B_{ij}\omega_j+\bar B_{ij}\bar\omega_j,
	\end{equation}
	Their periods \eqref{real_harmonic_basis} require 
	\begin{equation}
		A+\bar A=\mathbf 1,
		\qquad A\Gamma+\bar A\bar\Gamma=0,
		\qquad
		B+\bar B=0,
		\qquad B\Gamma+\bar B\bar\Gamma=\mathbf 1.
	\end{equation}
	Solving these linear equations yields
	\begin{equation}
		A=\frac12\big(\mathbf 1+iXY^{-1}\big),
		\qquad
		B=-\frac{i}{2}Y^{-1}.
	\end{equation}
	
	It remains to compute the norms of these forms. The relevant
	identities are particularly simple in complex coordinates. Locally,
	\begin{equation} \label{holomorphic-hodge-star}
		dz\wedge dz=0,
		\qquad
		dz\wedge d\bar z=-2i\,dx\wedge dy,
		\qquad
		*dz=-i\,dz,
		\qquad *d\bar z=i\,d\bar z.
	\end{equation}
	For two closed one-forms the Riemann bilinear identity expresses their wedge product in terms of their periods over $a_i,b_i$. Applying it to $\omega_i$ and $\bar\omega_j$ gives
	\begin{equation}
		\int_\Sigma\omega_i\wedge\bar\omega_j
		=\sum_{k=1}^g\left(
		\oint_{a_k}\omega_i\oint_{b_k}\bar\omega_j
		-\oint_{b_k}\omega_i\oint_{a_k}\bar\omega_j\right)
		=\bar\Gamma_{ij}-\Gamma_{ij}=-2iY_{ij}.
	\end{equation}
	Applied to two holomorphic forms, the same identity gives $\Gamma_{ij}=\Gamma_{ji}$, since $\omega_i\wedge\omega_j=0$. Together with $*\omega_i=-i\omega_i$ and $*\bar\omega_i=i\bar\omega_i$ that follows from \eqref{holomorphic-hodge-star}, this determines all the Hodge inner products. Substituting the coefficients $A$ and $B$,
	we obtain
	\begin{align}\label{hodge_matrices_period_matrix}
		\int_\Sigma\mathsf a_i\wedge *\mathsf a_j
		&=(Y+XY^{-1}X)_{ij},\notag\\
		\int_\Sigma\mathsf b_i\wedge *\mathsf b_j
		&=(Y^{-1})_{ij},\notag\\
		\int_\Sigma\mathsf a_i\wedge *\mathsf b_j
		&=-(XY^{-1})_{ij}.
	\end{align}
	
	For each choice of the periods in \eqref{boson_periods}, there is a unique harmonic one-form representative of that cohomology class,
	\begin{equation}\label{boson_classical_dphi}
		d\phi_{\rm cl}=2\pi R\left[
		\left(m_i+\frac{\beta_i}{2\pi}\right)\mathsf a_i
		+\left(n_i+\frac{\alpha_i}{2\pi}\right)\mathsf b_i\right].
	\end{equation}
	We decompose $\phi=\phi_{\rm cl}+\phi_{\rm qu}$, where the fluctuations $d\phi_{\rm qu}$ are an exact form. The cross term in the action vanishes by integration by parts because $d*d\phi_{\rm cl}=0$. Thus, for the action $S=\frac{1}{2}\int_\Sigma d\phi\wedge*d\phi$, the partition function is a product of a holonomy-independent quantum determinant and a sum over classical sectors. Writing $\boldsymbol M=\boldsymbol m+\boldsymbol\beta/(2\pi)$ and $\boldsymbol N=\boldsymbol n+\boldsymbol\alpha/(2\pi)$, the Hodge products above give, with $\lambda=2\pi R^2$ as in the main text,
	\begin{align}\label{classical_action_period_matrix}
		S_{\rm cl}
		&=\pi\lambda\left[
		\boldsymbol M^{\rm T}(Y+XY^{-1}X)\boldsymbol M
		+\boldsymbol N^{\rm T}Y^{-1}\boldsymbol N
		-2\boldsymbol M^{\rm T}XY^{-1}\boldsymbol N\right]\notag\\
		&=\pi\lambda\left[
		\boldsymbol M^{\rm T}Y\boldsymbol M
		+(\boldsymbol N-X\boldsymbol M)^{\rm T}
		Y^{-1}(\boldsymbol N-X\boldsymbol M)\right].
	\end{align}
	For the two-interval replica surface, the period matrix is purely imaginary, $\Gamma_n=iY_n$, and $X=0$. The two winding lattices therefore decouple,
	\begin{equation}\label{boson_classical_pure_imag}
		S_{{\rm cl}}=\pi\lambda\left[\left(\boldsymbol{m}+\frac{\boldsymbol{\beta}}{2\pi}\right)^{T}Y_{n}\left(\boldsymbol{m}+\frac{\boldsymbol{\beta}}{2\pi}\right)+\left(\boldsymbol{n}+\frac{\boldsymbol{\alpha}}{2\pi}\right)^{T}Y_{n}^{-1}\left(\boldsymbol{n}+\frac{\boldsymbol{\alpha}}{2\pi}\right)\right].
	\end{equation}
	The sum over $\boldsymbol n$ can be converted by Poisson resummation,
	\begin{align}\label{boson_poisson_resum}
		\sum_{\boldsymbol{n}\in\mathbb{Z}^{g}}e^{-\pi\lambda(\boldsymbol{n}+\frac{\boldsymbol{\alpha}}{2\pi})^{{\rm T}}Y_{n}^{-1}(\boldsymbol{n}+\frac{\boldsymbol{\alpha}}{2\pi})}=\frac{\sqrt{\det Y_{n}}}{\lambda^{g/2}}\sum_{\boldsymbol{k}\in\mathbb{Z}^{g}}e^{-\frac{\pi}{\lambda}\boldsymbol{k}^{{\rm T}}Y_{n}\boldsymbol{k}+i\boldsymbol{k}^{{\rm T}}\boldsymbol{\alpha}}.
	\end{align}
	The prefactor is independent of the holonomies, and hence cancels in the ratio between partition functions. We are left with the factorized classical sum
	\begin{align}\label{classical_factorized_sum}
		Z_{{\rm n}}(\boldsymbol{\alpha},\boldsymbol{\beta})&\propto\left[\sum_{\boldsymbol{m}\in\mathbb{Z}^{g}}e^{-\pi\lambda(\boldsymbol{m}+\frac{\boldsymbol{\beta}}{2\pi})^{{\rm T}}Y_{n}(\boldsymbol{m}+\frac{\boldsymbol{\beta}}{2\pi})}\right]\times\left[\sum_{\boldsymbol{k}\in\mathbb{Z}^{g}}e^{-\frac{\pi}{\lambda}\boldsymbol{k}^{{\rm T}}Y_{n}\boldsymbol{k}+i\boldsymbol{k}^{{\rm T}}\boldsymbol{\alpha}}\right]. \notag \\
		&\propto\Theta\!\begin{bmatrix}\tilde{\boldsymbol{\beta}}\\
			\boldsymbol{0}
		\end{bmatrix}(0|\lambda\Gamma_{n})\Theta\!\begin{bmatrix}\boldsymbol{0}\\
			\tilde{\boldsymbol{\alpha}}
		\end{bmatrix}(0|\Gamma_{n}/\lambda).
	\end{align}

	\bibliography{ref}\bibliographystyle{JHEP}

\end{document}